\documentclass[a4paper, fleqn, usenatbib]{mnras}

\usepackage[T1]{fontenc}
\usepackage{ae,aecompl}

\usepackage{graphicx}	
\usepackage{amsmath}	
\usepackage{amssymb}	
\usepackage{natbib}

\usepackage[usenames, dvipsnames]{color}
\usepackage[caption=false]{subfig}
\usepackage{hyperref}

\title[Compact galaxies quenched in clusters at $z<1$]{Compact star-forming galaxies preferentially quenched to become PSBs in \textit{z} < 1 clusters}

\author[M. Socolovsky et al.]
{Miguel Socolovsky,$^{1}$\thanks{E-mail: miguel.socolovsky@nottingham.ac.uk}
David T. Maltby,$^{1}$
Nina A. Hatch,$^{1}$
Omar Almaini,$^{1}$
\newauthor Vivienne Wild,$^{2}$
William G. Hartley,$^{3}$
Chris Simpson,$^{4}$
Kate Rowlands$^{5}$
\\
$^{1}$School of Physics and Astronomy, University of Nottingham, Nottingham NG7 2RD, UK\\
$^{2}$School of Physics and Astronomy, University of St Andrews, North Haugh, St Andrews, KY16 9SS, UK\\
$^{3}$Department of Physics and Astronomy, University College London, 3rd Floor, 132 Hampstead Road, London NW1 2PS, UK\\
$^{4}$Gemini Observatory, Northern Operations Center, 670 N.~A`ohoku Place, Hilo, HI 96720-2700, USA\\
$^{5}$Department of Physics and Astronomy, Johns Hopkins University, Bloomberg Center, 3400 N. Charles St., Baltimore, MD 21218, USA
}

\date{Accepted 2018 October 18. Received 2018 October 12; in original form 2018 August 6}

\pubyear{2018}

\begin{document}
\label{firstpage}
\pagerange{\pageref{firstpage}--\pageref{lastpage}}
\maketitle

\begin{abstract}
We analyse the structure of galaxies with high specific star formation rate (SSFR) in cluster and field
environments in the redshift range $0.5 < z < 1.0$.  Recent studies have shown that these galaxies are
strongly depleted in dense environments due to rapid environmental quenching, giving rise to
post-starburst galaxies (PSBs). 
We use effective radii and S\'ersic indices as tracers of galaxy structure, determined using imaging from 
the UKIDSS Ultra Deep Survey (UDS). We find that the high-SSFR galaxies that survive into the cluster
environment have, on average, larger effective radii than those in the field. We suggest that this trend is likely to be driven by the most compact star-forming galaxies being preferentially quenched in dense environments. We also show that the PSBs in clusters have stellar masses and effective radii that are similar to 
the missing compact star-forming population, suggesting that these PSBs are the result of size-dependent quenching. We propose that both 
strong stellar feedback and the stripping of the extended halo act 
together to preferentially and rapidly quench the compact and low-mass star-forming 
systems in clusters to produce PSBs. We test this scenario using the stacked spectra of 124 
high-SSFR galaxies, showing that more compact galaxies are more likely to host 
outflows. We conclude that a combination of environmental and secular processes is the most likely 
explanation for the appearance of PSBs in galaxy clusters.

\end{abstract}




\section{Introduction}

In the local Universe the most massive galaxies are passive and elliptical, while low-mass galaxies tend to
be star-forming and have disc-dominated morphologies.  Several studies at high redshift show that massive
passive galaxies are already in place at $z = 2$ \citep{vanderwel2008a, bamford2009, baldry2012}.  In
contrast, the observed number of quenched low-mass galaxies increases towards the present day
\citep{drory2009, baldry2012, moutard2016b, moutard2018}.  This downsizing in the passive population is
generally associated with environmental quenching, and has been measured up to $z\sim1$ \citep{muzzin2013,
tomczak2014, socolovsky2018}.  These evolutionary trends are consistent with galaxies in high-density
environments being more likely to be passive regardless of their stellar mass \citep{balogh1997, baldry2006,
yang2009, peng2010, hartley2013}. 

There are various proposed mechanisms to explain environmental quenching, but consensus has not yet been reached 
on which are the dominant processes. Interactions with the intra-cluster medium, such as strangulation 
\citep{larson1980} or ram-pressure stripping \citep{gunn1972}, are some of the preferred mechanisms to explain 
how star-forming galaxies are deprived of their gas reservoirs and subsequently quench. Alternative mechanisms invoke 
galaxy-galaxy interactions, such as mergers, harassment or tidal stripping \citep{moore1996, toomre1972, faber1973}, 
which are examples of other processes typically associated with high and intermediate-density environments.

Galaxy structure provides a window into the evolutionary history of galaxies, so studying the structure of 
different galaxy populations can help to disentangle the driving quenching mechanisms. 
Gravitational interactions (including major mergers) may induce the migration of gas and stars towards the  
galaxy centre, producing more compact and concentrated light profiles. This contrasts with the faded discs 
generated when the gas is ram-pressure stripped via interaction with the intra-cluster medium (ICM). 
The environmental dependence of the galaxy stellar-mass--size relation for early-type galaxies 
has been extensively studied in the past. \citet{cooper2012} and \citet{lani2013} used local density as 
a tracer of environment at $z>1$, and found that red sequence galaxies at fixed stellar mass 
present larger radii in high density environments. A different type of study, i.e. comparing cluster 
and field galaxies at $z=1.6$, also showed that early-type galaxies are larger in the cluster environment 
\citep{papovich2012}. In contrast, at lower redshift \citet{kelkar2015} found no significant difference between cluster and field 
galaxies at $z\sim0.6$. They concluded that the size evolution in the field might have caught up with the 
cluster, erasing the observed differences at higher redshifts.

Previous studies seem to agree that the growth in size of passive galaxies in dense 
environments is driven by dry merging and tidal interactions \citet[e.g.][]{cooper2012, lani2013}. However, not much work has been done on how the
mass--size relation of star-forming galaxies is affected by environment at high redshift. In the low-redshift 
Universe, some authors have found that late-type galaxies are larger in the field than in galaxy clusters 
\citep{maltby2010, cebrian2014}. In this paper we extend the study of the mass--size relation of star-forming 
galaxies in different environments to $z=1$.

The most depleted population in clusters are those with low stellar masses and high specific star formation 
rates (SSFRs). In \citet{socolovsky2018} we showed that these high SSFR galaxies are likely to evolve 
into low-mass post-starbursts (PSBs) in dense environments at $0.5<z<1.0$. By studying the stellar-mass function, 
we found a significant excess of low-mass PSBs in clusters. The mass distribution of these recently 
quenched galaxies in clusters is only comparable to the mass function of young, high-SSFR galaxies 
(SF1 galaxies) in the field. Therefore, SF1 galaxies are likely to be the progenitors of PSBs in dense environments, 
where they experience rapid environmental quenching. In this paper we investigate the quenching mechanisms by 
studying the stellar-mass--size relation of PSBs and their progenitors, SF1 galaxies.

The structure of the paper is as follows.
In Section~\ref{sec:data} we present our data, the classification method and a description of how structural 
parameters are measured from ground-based imaging. We also present a brief description of the cluster-finding 
algorithm, developed in \citet{socolovsky2018}. We present our results in Section~\ref{sec:results} 
and discuss their possible implications in Section~\ref{sec:discussion}. Finally, our conclusions are listed 
in Section~\ref{sec:conclusions}. Throughout this paper we use AB magnitudes measured using 2'' apertures, and assume a $\rm\Lambda CDM$ cosmology 
with the following parameters: $\Omega_M=0.3$, $\Omega_\Lambda=0.7$ and $H_0=70~\rm kms^{-1}Mpc^{-1}$.

\section{Data sets and galaxy classification}
\label{sec:data}

\subsection{Galaxy catalogue}
\label{sec:catalogue}

The galaxy catalogue we use is based on the 8th data release of the UDS (Almaini et al., in prep), with coverage in the near-infrared to $5\sigma$ depths measured in 2'' apertures of $J=24.9$, $H=24.2$ and $K=24.6$. 
Deep optical imaging is also available from the Subaru \textit{XMM--Newton} Deep Survey (SXDS; \citealt{furusawa2008, ueda2008}), with $5\sigma$ depths of $B=27.6$, $V=27.2$, $R=27.0$, $i'=27.0$ and $z'=26.0$.
Deep $U$-band  imaging is provided by the Canada-France-Hawaii Telescope (CFHT), to a depth of $U=26.75$ to $5\sigma$ and the $Spitzer$ Legacy Program (SpUDS) provides $\left[3.6\right]=24.2$ and $\left[4.5\right]=24.0$ at $5\sigma$. The combined area, after masking, covers $\sim0.62$~square degrees.

The catalogue is limited to $K$<24.3 which ensures a $95\%$ completeness \citep{hartley2013}. The missing $5\%$ corresponds mainly to low surface brightness galaxies. This leads to a catalogue with 23,398 galaxies at $0.5<z<1.0$. Stars are identified and removed according to the method described in \citet{simpson2013}.

\subsection{Photometric redshifts and stellar masses}
\label{sec:photozs}

We use the photometric redshifts derived by \citet{simpson2013} based on the UDS DR8 photometry. They use the {\sc eazy} photometric-redshift code \citep{brammer2008} to fit template spectra to the $U$, $B$, $V$, $R$, $i'$, $z'$, $J$, $H$, $K$, $3.6\mu \text{m}$ and $4.5\rm\mu m$ photometry. Approximately 1500 spectroscopic redshifts from the UDSz (ESO Large Programme, Almaini et al., in prep) and 3500 from the literature \citep{simpson2012} were used to test these photometric redshifts. The measured median absolute deviation is \mbox{$\sigma_{\text{NMAD}}\left(\Delta z/(1+z)\right)\sim0.023$} up to $z=1.0$, with $<4\%$ outliers, defined as sources with $\Delta z/(1+z)>5\sigma_{\rm NMAD}$, once AGN are removed. Galaxy redshifts were fixed to the spectroscopic values, where available, otherwise photometric redshifts were assumed.

The stellar masses are also calculated in \citet{simpson2013} by fitting a grid of synthetic spectral energy distributions (SEDs) to the 11-band photometry, built using \citet{bruzual2003} stellar population synthesis models and using the \citet{chabrier2003} initial mass function.

\subsection{Galaxy classification and SSFRs}
\label{sec:classification}

We use the galaxy classification described in \citet{wild2016} which is based on a Principal Component 
Analysis (PCA) outlined in \citet{wild2014}. See these two papers for a more detailed description 
of the method. We provide here a brief overview of the key features as well as describe the galaxy 
classes we use.

Three eigenspectra are determined using a PCA on a grid of 44,000 model SEDs, using the stellar population 
synthesis models from \citet{bruzual2003}. The coefficients that quantify the contribution 
of each of these eigenspectra in order to reproduce one galaxy SED are named `supercolours'. The 
first supercolour, $SC1$, modifies the red-blue slope and traces the $R$-band weighted mean stellar 
age or SSFR. Supercolour $SC2$ alters the strength of the Balmer break 
and correlates with the fraction of the stellar mass formed in bursts during the last billion years, 
and also traces metallicity. Supercolour $SC3$ affects the shape of the SED around $4000$~\r{A} and 
is used to break the degeneracy between metallicity and the fraction of the stellar mass formed during the 
last Gyr.

The PCA allows us to derive other physical properties of galaxies, which are obtained from the grid of models described above.
Hence, the SSFRs we use in this work are computed using the supercolour method \citep[see][]{wild2016}. Although the PCA can provide stellar masses, we choose to use the masses from \citet{simpson2013} derived using SED fitting. These masses are better constrained as all 11 photometric bands are used, rather than the 8 used in the PCA \citep[see][for more details]{wild2016}.

The galaxy classification is based on the position of galaxies in SC-SC diagrams (typically SC1-SC2 and SC1-SC3) 
and the boundaries between populations are determined empirically using model SEDs and spectroscopy (see 
\citealt{wild2014}). This method divides the population into star-forming (SF), passive (PAS), post-starburst 
(PSB), metal-poor and dusty galaxies (the last two are excluded from our sample). Additionally, the SF 
population is also divided into three subpopulations of decreasing SSFR: SF1, SF2 and SF3. 
The method has also been spectroscopically confirmed \citep{wild2014, maltby2016}.

The PCA is applied to a catalogue with a magnitude limit of $K$<24. This catalogue is slightly more conservative than the one described 
in Section~\ref{sec:catalogue}, which reduces the noise in the supercolour determination (see \citealt{wild2016}). This leads to $11,\!625$~SF1, $3,\!486$~SF2, 
$2,\!055$~SF3, $2,\!206$~PAS and $418$~PSBs in the redshift range $0.5 < z < 1.0$. We compute $90\%$ mass 
completeness limits for each galaxy type using the method of \citet{pozzetti2010}. The values of 
$\log M_{\rm lim}/M_\odot$ to ensure a 95\% mass completeness at $z=1$ are 9.0 for SF, 9.5 for PAS and 9.3 for 
PSB galaxy populations.

\subsection{Cluster and field samples}
\label{sec:clusters}

The cluster and field samples are drawn from \citet{socolovsky2018}. The classification 
method is based on a friends-of-friends algorithm which runs on the $K$-band galaxy catalogue 
of the UDS. This algorithm depends on three parameters which are tested and optimized 
by running the algorithm on a mock catalogue which includes simulated galaxy clusters (we 
refer the reader to \citealt{socolovsky2018} for details).

The sample, consisting of 37 galaxy overdensities at $0.5<z<1.0$, is likely to 
be dominated by group-like structures ($\sigma_{v}=300$--$500\rm\,km\,s^{-1}$) combined 
with more massive galaxy clusters. For the purposes of this study, henceforth we refer to these 
overdensities collectively as ``clusters''. 
A threshold of at least 20 detected members is applied to ensure a high 
signal-to-noise ($S/N$; see \citealt{socolovsky2018}). Every galaxy located within 1~Mpc 
from the projected centre of mass and $2.5\sigma_z$ ($\sigma_z=0.023(1+z)$) from the median 
redshift of the measured structure was included in the cluster sample to ensure membership completeness. 
These criteria are applied to both galaxies with and without spectroscopic redshifts for consistency.
This cluster sample, therefore, contains contaminants from the field.
The field sample is constructed using all the 
galaxies in the UDS field that were not associated with an overdensity, while 
forced to follow the redshift distribution of the cluster sample. In total the 
sample has 2,210 cluster galaxies and 13,837 in the field between $0.5<z<1.0$.

\subsection{Galaxy size and S\'ersic index from UDS DR11}
\label{sec:structural_param}

Structural parameters (i.e. effective radius, $R_e$ and S\'ersic index, $n$) were determined 
using the $K$-band image from the UDS DR11 ($J=25.6$, $H=25.1$, $K=25.3$; $5\sigma$, AB). The 
software employed was {\sc GALAPAGOS} \citep{barden2012}, which makes use of {\sc GALFIT} 
\citep{peng2002} in order to fit a S\'ersic light profile \citep{sersic1968} to each galaxy 
in the UDS. We refer the reader to \citet{almaini2017} for further details.

We rejected poor fits ($\chi^2_\nu>100$) which corresponds to $1.7\%$ of our sample. 
Similarly, we rejected $\sim 7\%$ of galaxies where {\sc GALFIT} did not converge to one solution. 
Most of these rejections correspond to objects with low surface brightness and near masked regions.
The rejection rate was similar for star-forming, passive and PSB galaxies; and for cluster and field galaxies. 

\begin{figure}
 	\begin{center}
 		\includegraphics[width=0.5\textwidth]{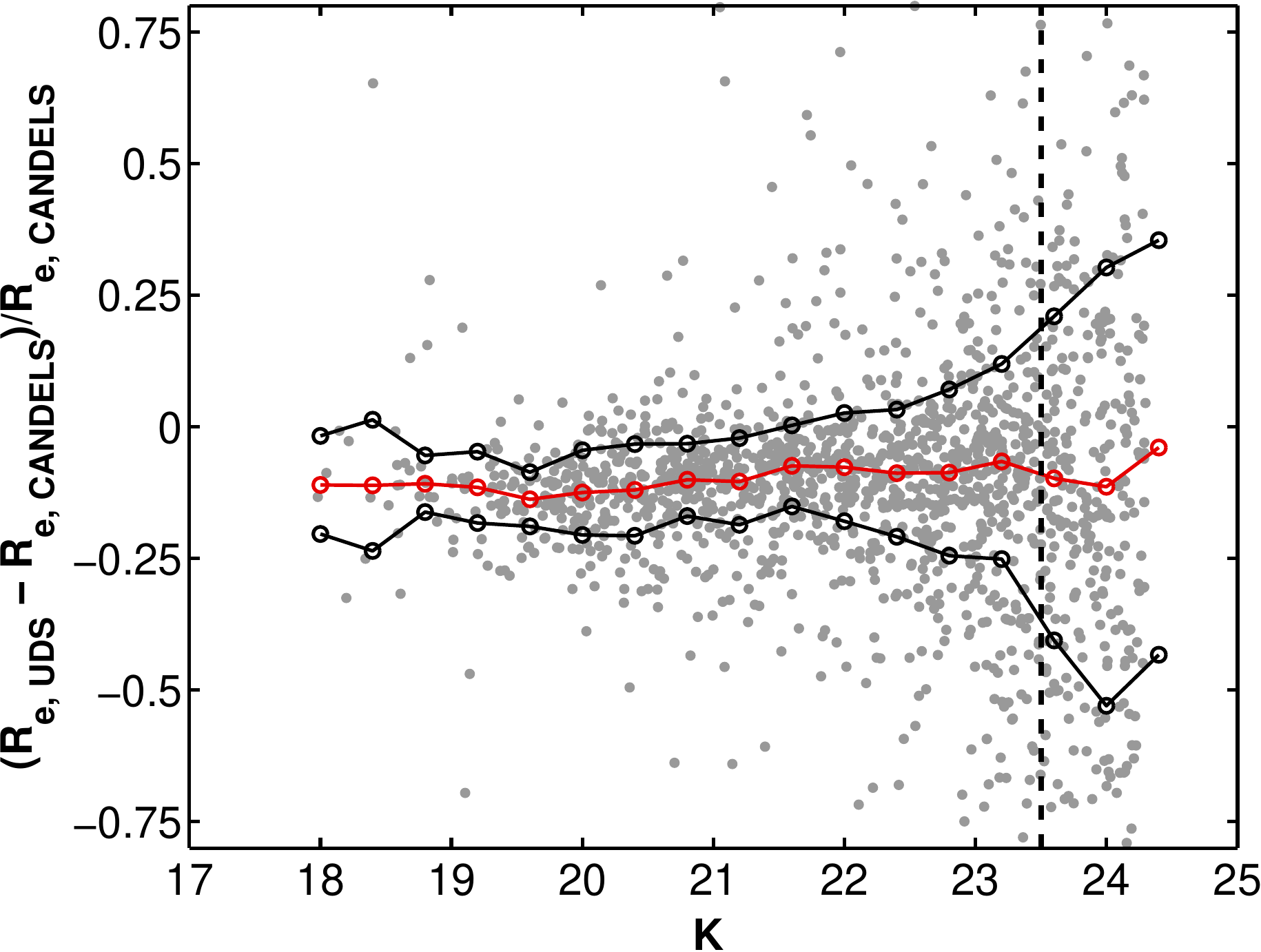}
		\vspace{-10pt}
 		\caption{Relative difference between the effective-radii measured from ground-based UDS DR11 $K$-band imaging and {\it HST} CANDELS $H$-band imaging as a function of $K$-magnitude ($0.5<z<1.0$). The median values and median absolute deviations are displayed as red and black circles, respectively. Ground-based sizes are systematically 10\% smaller than the ones measures from space. This is due to both the lower background noise in space-based images and the expected variation between filters \citep{kelvin2012}. We choose a magnitude limit of $K=23.5$ (vertical line), which corresponds to a 25\% scatter, to only select reliable effective radii.}
 		\label{fig:candels_re_vs_K}
 	\end{center}
 \end{figure}

In Fig.~\ref{fig:candels_re_vs_K} we compare our $K$-band sizes with those obtained using the $H$-band from 
the overlapping {\it Hubble Space Telescope} ({\it HST}) CANDELS survey \citep{vanderwel2012}, which covers 
$\sim7\%$ of the UDS field. Space- and ground-based effective radii are found to be in good agreement. 
We find that ground-based sizes are systematically $10\%$ smaller than the space-based ones, which 
is consistent with the expected variation across wavelengths \citep{kelvin2012}. We impose a $K$-band 
cut of $K=23.5$ (vertical line in Fig.~\ref{fig:candels_re_vs_K}) to reject faint 
galaxies with unreliable $R_{\rm e}$ values. This flux limit corresponds to a 25\% scatter in $\delta R_{\rm e}/R_{\rm e}$, 
estimated using the normalized median absolute deviation and rejects 11.5\% of the total sample. 

After applying these quality cuts we are left with a sample of 
5421 (1453) SF, 1146 (307) PAS and 95 (26) PSB field (cluster) galaxies.

\section{Results}
\label{sec:results}

\subsection{Galaxy size as a function of specific star formation rate}
\label{sec:size-SSFR}

 \begin{figure}
 	\begin{center}
 		\includegraphics[width=0.45\textwidth]{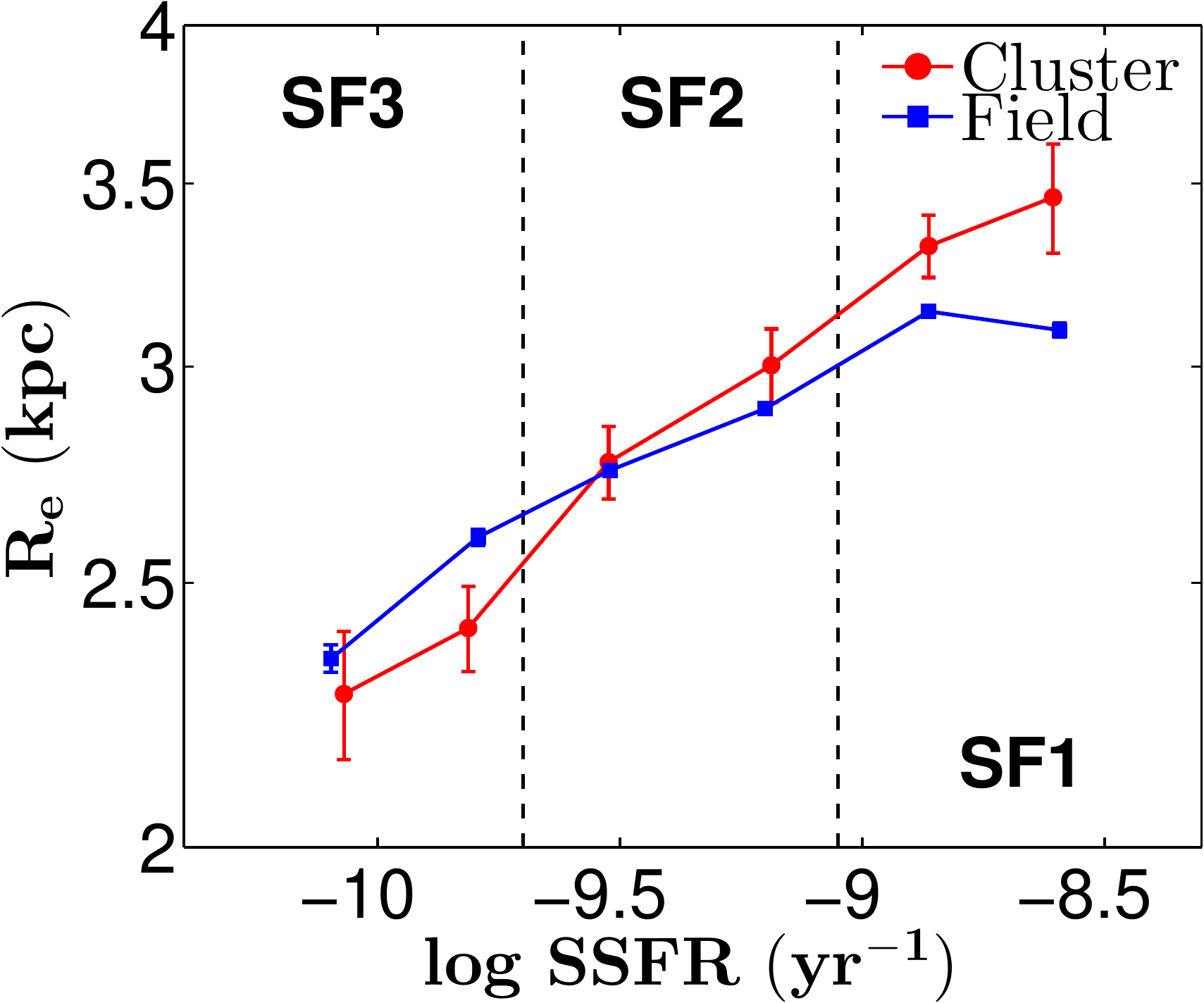}
		\vspace{-8pt}
 		\caption{Median effective radius of star-forming galaxies as a function of SSFR at $0.5<z<1.0$. The red and blue lines correspond to cluster and field environments, respectively. The data points are centred on the median SSFR in each bin and the $1\sigma$ confidence error bars are estimated using bootstrapping. The vertical dashed lines delimit the regions typically occupied by the different star forming populations: SF1, SF2 and SF3, in order of decreasing mean SSFR. The 
		SF1 galaxies (with the highest SSFR) are found to be, on average, larger in the cluster environment than in the field.}
 		\label{fig:SF_SSFR-RE}
 	\end{center}
 \end{figure}
 \begin{figure}
 	\begin{center}
 		\includegraphics[width=0.45\textwidth]{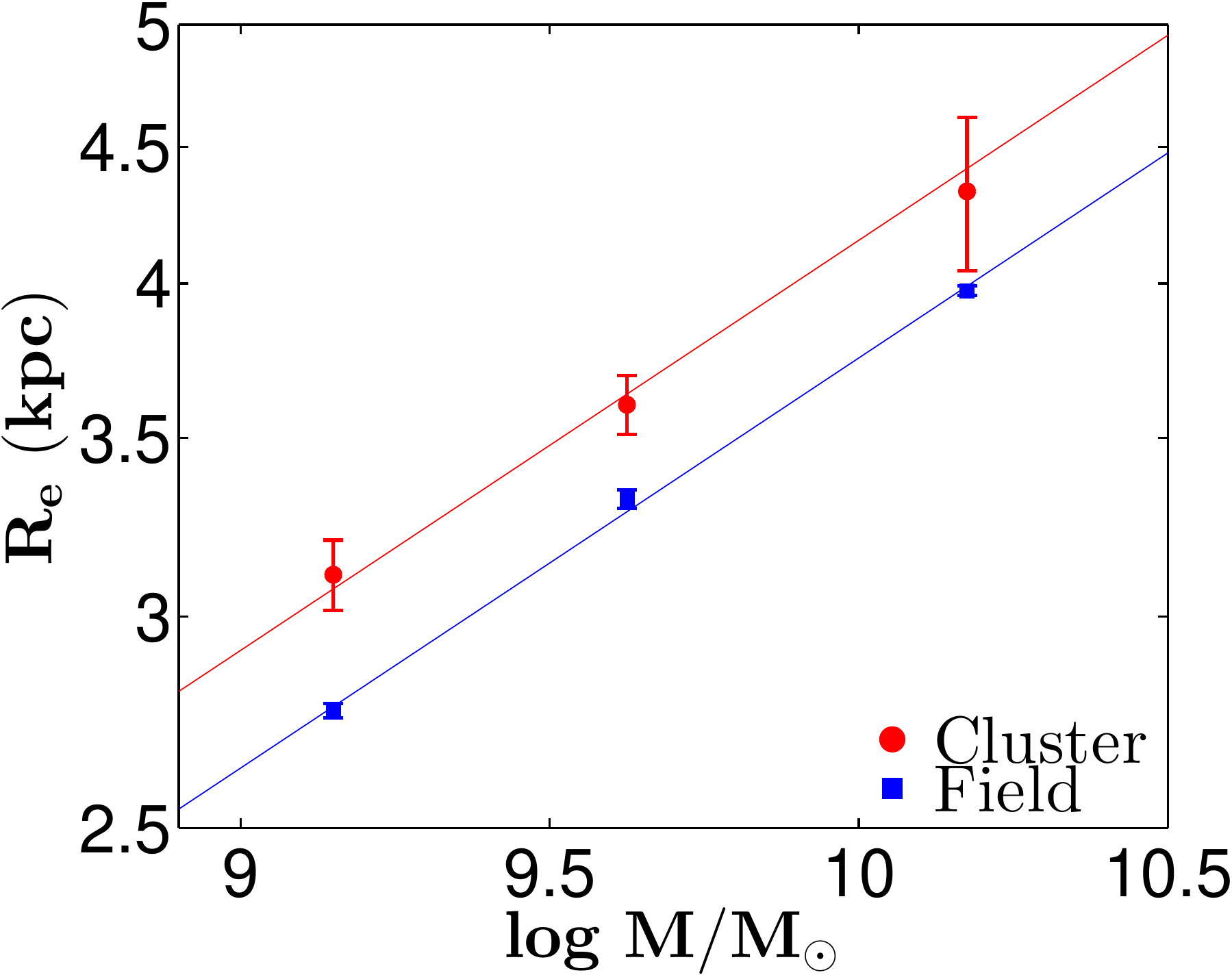}
		\vspace{-8pt}
 		\caption{Stellar-mass--size relations of cluster (red) and field (blue) high-SSFR galaxies (SF1 galaxies) at $0.5<z<1.0$. The errorbars correspond to the $1\sigma$ confidence intervals estimated using bootstrapping. We fit a linear model to each mass--size relation (solid lines) in order to compare them. We find that cluster SF1 galaxies are systematically 9\% larger than their field counterparts at all masses.}
 		\label{fig:SF1_mass-size}
 	\end{center}
\end{figure}

In Fig.~\ref{fig:SF_SSFR-RE} we show the dependence of the median 
$R_e$ on SSFR for the star-forming galaxies in our sample. We 
find that $R_e$ increases approximately linearly 
with $\log SSFR$. When we split galaxies 
by environment, we observe that cluster galaxies with high SSFRs 
have larger median $R_e$ than their field counterparts. 
The vertical dashed lines in Fig.~\ref{fig:SF_SSFR-RE} correspond to 
the approximate boundaries between the three star-forming 
subpopulations described in Section~\ref{sec:classification}, which 
correlate well with SSFR. Thus, most of the galaxies with 
$\text{SSFR}>10^{-9.0}~\text{yr}^{-1}$ belong to the population of 
young star-forming galaxies, i.e. SF1. 

In Fig.~\ref{fig:SF1_mass-size} we show the stellar-mass--size relation 
for SF1 galaxies as a function of environment. The 
values correspond to the median galaxy size in each mass bin, and the errorbars 
represent the error on the median, estimated using a bootstrapping technique. The 
lower mass limit corresponds to the mass completeness limit ($10^{9.0}\rm M_\odot$). 
The stellar-mass--size relation does not extend beyond $10^{10.5}\rm M_\odot$ 
because there are no SF1 galaxies with higher masses in our survey. 
As expected, in both environments galaxies increase in size for increasing stellar mass 
\citep{shen2003, vanderwel2014}. However, it is evident that the remaining SF1 galaxies 
in clusters (that are not yet significantly affected by environmental quenching) are on 
average larger than in the field at all stellar masses. 
We fit a linear model with fixed slope ($\log R_{\rm e}/\log M_* = 0.154$) to the data from both 
environments and compare the intercepts to quantify the level of agreement. The intercept 
value in the cluster environment is $\log R_{\rm e} (M_*=10^{9.5}M_\odot) = 0.542\pm0.013$, 
in contrast with a field value of $\log R_{\rm e} (M_*=10^{9.5}M_\odot) = 0.498\pm0.002$. 
This represents a $3.4\sigma$ discrepancy between the cluster and field environments. 
We note, however, that our cluster sample contains contaminants from the field, which dilutes 
the differences between environments. Hence, the difference measured here is likely to be a lower 
limit and the real level of significance may be much higher. 

In Figs.~\ref{fig:SF_SSFR-RE} \& \ref{fig:SF1_mass-size} we have shown that the SF1 galaxies that survive 
in the cluster are on average larger than the general SF1 population in the field. 
This is unlikely to be driven by an increase in the SSFR of cluster galaxies as a result of an interaction with the cluster environment. This is because SF1 galaxies are the largest population in size ($R_{\rm e}$), on average (see Fig.~\ref{fig:SF_SSFR-RE}), so increasing the SSFR of SF2 or SF3 galaxies to become SF1 galaxies would decrease the median size rather than increase it.
From this, we conclude that dense environments affect the mass--size relation of young, highly star-forming galaxies.

\subsection{A lack of compact star-forming galaxies in galaxy clusters}
\label{sec:largeSF1}

 \begin{figure*}
 	\begin{center}
 		\includegraphics[width=0.99\textwidth]{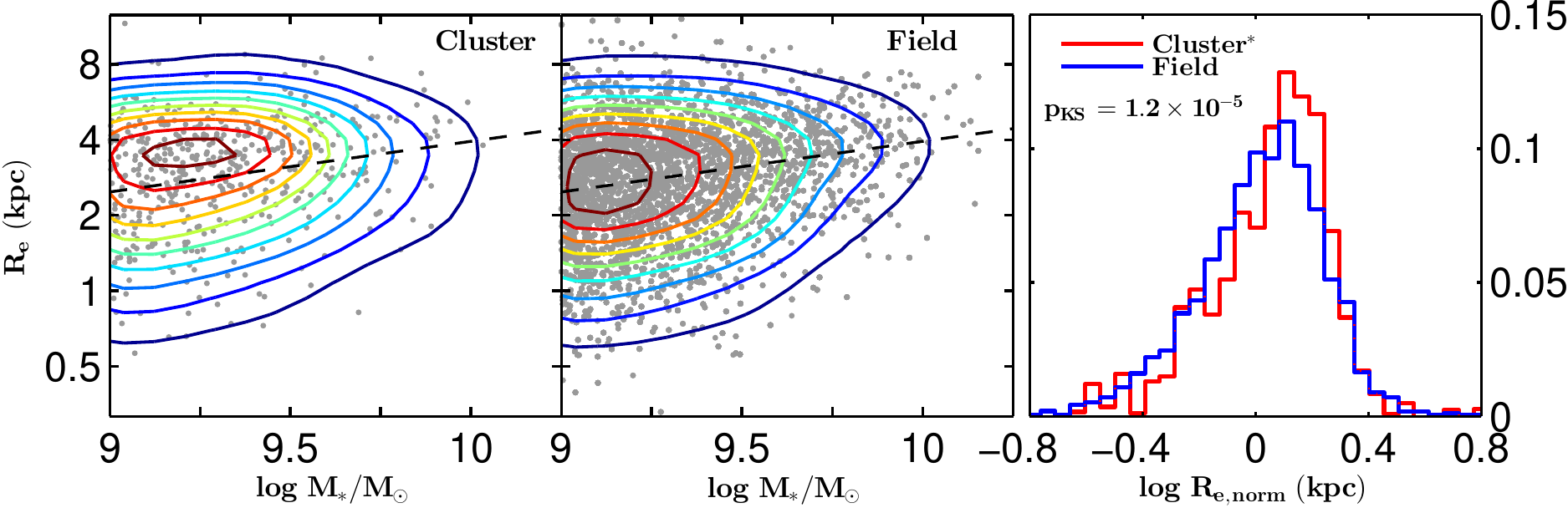}
		\vspace{-8pt}
 		\caption{The stellar-mass--size relation for SF1 galaxies in the UDS at $0.5<z<1.0$. The grey dots represent the position of individual galaxies across the mass--size plane. The contours show the number of galaxies per unit area on the diagram and normalized by the comoving volume of the field. The left and central panels correspond to cluster and field galaxies, respectively. The black dashed line in the first two panels corresponds to the best-fitting linear model to the field population. The right-hand panel represents the distribution of the variable ${R}_{\rm e, norm}$ or the ratio between the effective radius of a galaxy and the value predicted by the best-fit model to the field data. The red line corresponds to the cluster and the blue to the field populations (note the asterisk next to the cluster label indicating it is the background-subtracted cluster sample). The field and cluster distributions are significantly different, according to a KS test ($p_{\rm KS}$ is quoted on the top left corner), with the cluster SF1 galaxies being, on average, larger than in the field.}
 		\label{fig:subtracted mass-size}
 	\end{center}
 \end{figure*}

\begin{figure*}
 	\begin{center}
 		\includegraphics[width=0.99\textwidth]{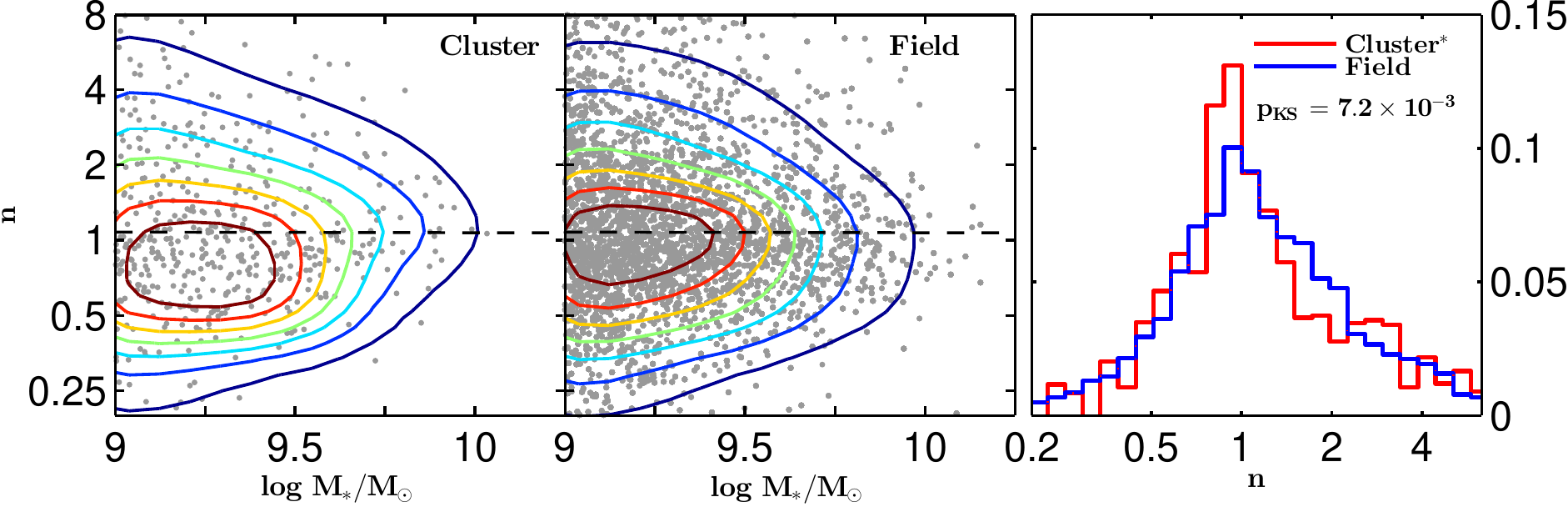}
		\vspace{-8pt}
 		\caption{The stellar-mass--S\'ersic index relation for SF1 galaxies in the UDS at $0.5<z<1.0$. The grey dots represent the position of individual galaxies across the mass--S\'ersic index plane. The contours show the number of galaxies per unit area on the diagram and normalized by the comoving volume of the field. The black dashed line in the first two panels corresponds to the best-fitting linear model to the field population. The left and central panels correspond to cluster and field galaxies, respectively. The right-hand panel shows the distributions of $n$ in clusters (red) and in the field (blue; the asterisk next to the cluster label indicates that it is the background-subtracted sample). We observe that cluster galaxies tend to have lower $n$ values with respect to the field.}
 		\label{fig:subtracted mass-sersic}
 	\end{center}
 \end{figure*}

In this section, we examine the distributions of galaxy size and S\'ersic index as a function of environment for our SF1 galaxies. 
We look first at the distribution of galaxies across the mass--size and mass--S\'ersic 
index planes (left and central panels of Figs.~\ref{fig:subtracted mass-size} and 
\ref{fig:subtracted mass-sersic}). The first two panels on the left in Fig.~\ref{fig:subtracted mass-size} 
show the distribution of SF1 galaxies on the stellar-mass--size plane, the left panel corresponds 
to clusters and the central one to the field. The straight line in both panels corresponds 
to a linear fit to the mass--size relation in the field, which we use as a reference, 
\begin{equation}
\label{eq:MSRF}
\log R_{\rm e} = 0.202 \log M_* - 1.426.
\end{equation}
We observe that the cluster and field distributions are notably different. The cluster 
distribution peaks above the field mass--size relation, indicating larger sizes at 
the same stellar mass. In Fig.~\ref{fig:subtracted mass-sersic} we look at the distribution 
of SF1 galaxies across the mass--S\'ersic index plane. The dashed line corresponds to a linear fit to 
the field mass--$n$ relation, which is consistent with $n=1$. 
We find that the cluster distribution (left) peaks at lower values of $n$ than in the 
field (centre).

In the right-hand panels of Figs.~\ref{fig:subtracted mass-size} and \ref{fig:subtracted mass-sersic} 
we compare the cluster and field distributions of $n$ and normalized $R_{\rm e}$, i.e. removing the 
mass dependence of the field sample (eq.~\ref{eq:MSRF}). To remove the contaminants from the cluster 
sample we statistically subtract the contribution due to field galaxies that are erroneously included 
in the cluster sample.

The distributions of $R_{\rm e, norm}$ and $n$ are normalized to unity to allow direct comparison of their shape.
We see that the distributions of $R_{\rm e, norm}$ in high and low-density environments are significantly different ($p_{\rm KS} = 1.2\times 10^{-5}$).
As suggested from the previous results, the cluster SF1 population is skewed towards higher $R_{\rm e, norm}$ values, as compared to the field.
This galaxy population is known to be strongly depleted in dense environments \citep{socolovsky2018}. 
This trend is likely to be produced by the preferential quenching of the compact SF1 galaxies.
We also find a moderate but significant difference in the distribution of $n$ between cluster and field SF1 galaxies ($p_{\rm KS} = 7.2\times 10^{-3}$). Cluster SF1 galaxies seem to have a narrower distribution around $n=1$, while they have slightly higher $n$ in the field. Thus, SF1 galaxies with slightly higher $n$ might also be preferentially quenched in clusters.

In summary, we find that the remaining SF1 galaxies in dense environments are on average larger, potentially
because the compact SF1 galaxies are preferentially missing in dense environments.  These compact SF1
galaxies also have higher S\'ersic indices.

\subsection{The cluster post-starburst mass--size relation}

In \cite{socolovsky2018}, we found that PSB galaxies are the descendants of SF1 galaxies in clusters at
$0.5 < z < 1.0$.  We found that the stellar mass function of cluster PSBs present a very distinctive steep
low-mass slope.  Such a steep slope is only matched by the SF1 mass function.  This implies that the only
possible progenitors to PSBs are SF1 galaxies.  To build on this result, we analyse the mass--size relation
of PSBs to gain insight into the potential transformations that SF1 galaxies undergo as they quench in
clusters.

\begin{figure}
 	\begin{center}
 		\includegraphics[width=0.45\textwidth]{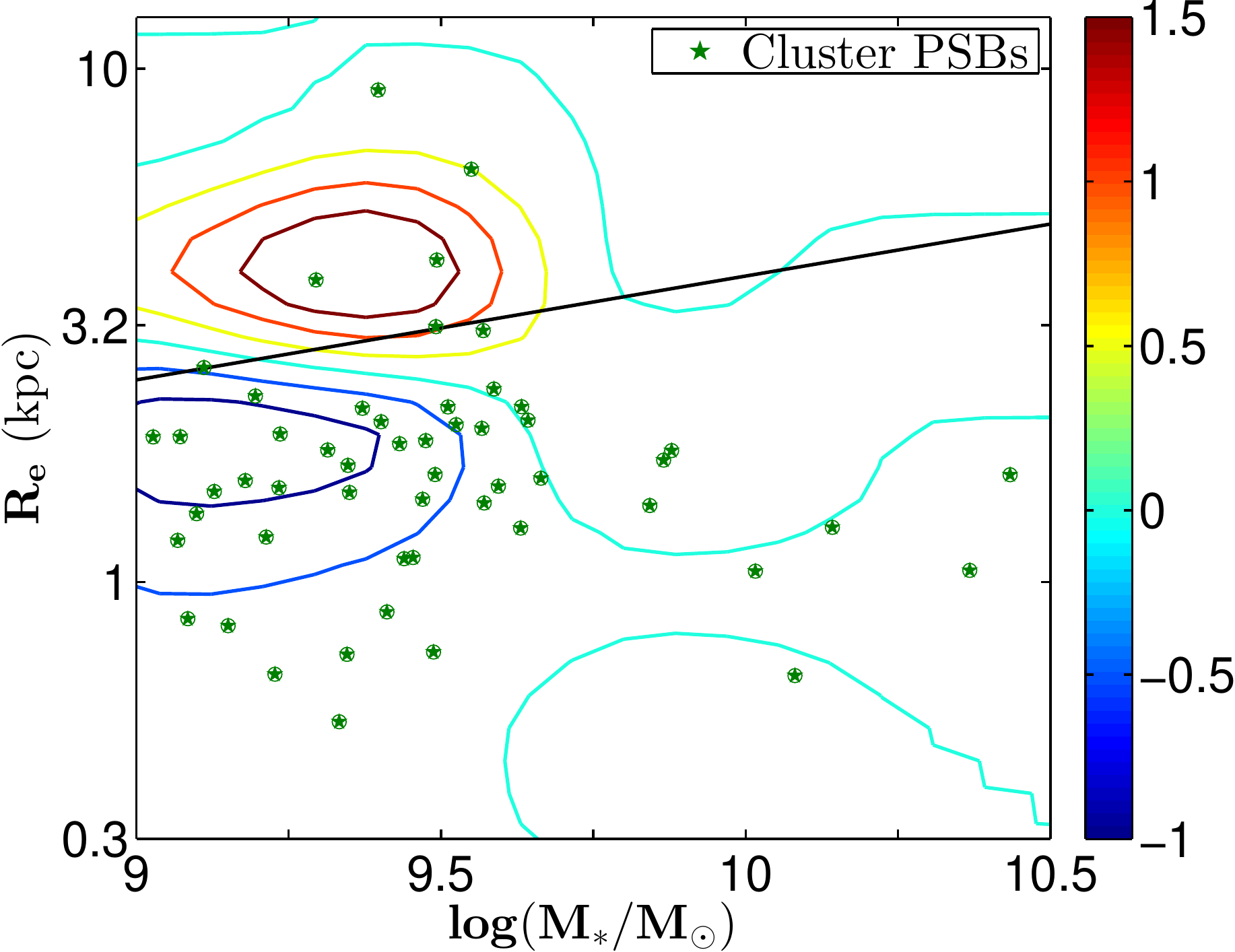}
		\vspace{-8pt}
 		\caption{Comparison of mass--size relations of SF1 galaxies and PSBs in the redshift range $0.5<z<1.0$. 
			 The contours represent the differential distribution of SF1 galaxies (i.e. the cluster minus the field distributions). The green stars show 
			 the location of cluster PSB galaxies. The solid line represents the mass--size relation of field SF1 galaxies for comparison. Cluster PSBs are located in the regions of the mass--size relation where SF1 galaxies are depleted in clusters with respect to the field, i.e. below the solid black line.}
 		\label{fig:PSB-Rdiff}
 	\end{center}
 \end{figure}

\begin{figure}
 	\begin{center}
 		\includegraphics[width=0.44\textwidth]{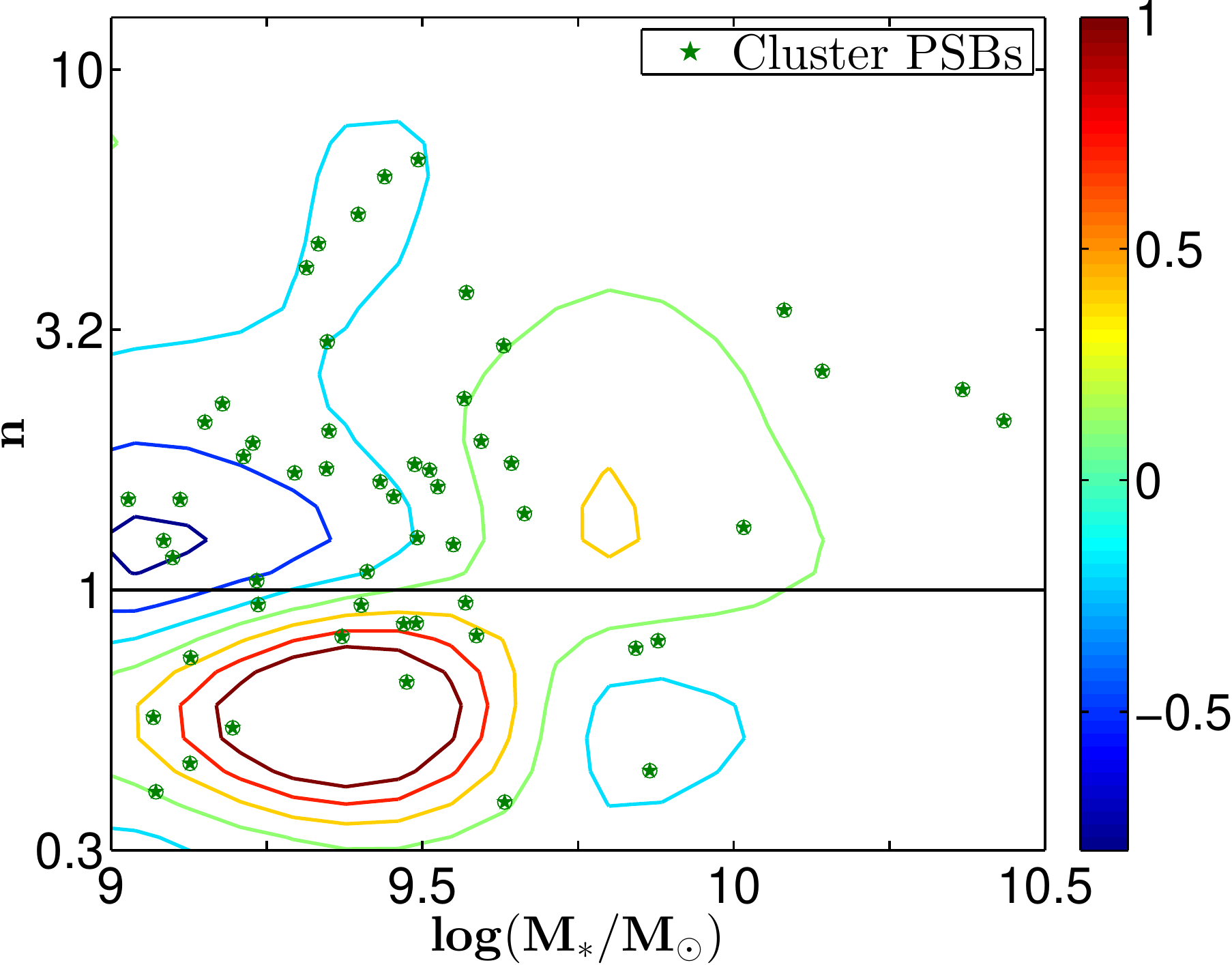}
		\vspace{-8pt}
 		\caption{Comparison of mass--S\'ersic index relations of SF1 galaxies and PSBs in the redshift range $0.5<z<1.0$. 
			 The contours represent the differential distribution of SF1 galaxies (i.e. the cluster minus the field distributions). The green stars show 
			 the location of cluster PSB galaxies. The solid line represents the mass--S\'ersic index relation of field SF1 galaxies for comparison. As in the case of the stellar mass--size relation, cluster PSBs are located in the regions where SF1 galaxies are depleted in clusters with respect to the field, in this case above the black line.}
 		\label{fig:PSB-ndiff}
 	\end{center}
 \end{figure}

Figs.~\ref{fig:PSB-Rdiff} and \ref{fig:PSB-ndiff} show the difference between the cluster and the field SF1
galaxy distributions across the mass--size and mass--S\'ersic index planes as colour contours.  Both cluster
and field distributions are normalized to unity to high-light their differences.  Superimposed on the
contours, the best-fitting line to the field SF1 mass--size (eq.~\ref{eq:MSRF}) and mass--$n$ relations are presented, 
to aid the comparison with Figs.~\ref{fig:subtracted mass-size} and
\ref{fig:subtracted mass-sersic}.

Although PSBs in clusters are generally more compact than the average size of the SF1 galaxies, we find that
the distribution of the cluster PSBs in the mass--size (Fig.~\ref{fig:PSB-Rdiff}) and mass--$n$
(Fig.~\ref{fig:PSB-ndiff}) planes coincides with the region where cluster SF1 galaxies are missing with
respect to the field, i.e.\ $48/54$ of the PSBs are found below the SF1 mass--size relation of the field.
Additionally, cluster PSBs typically have $n\sim1.5$, which indicates that they partially maintain a
disc-like nature. Furthermore, cluster PSBs and compact field SF1 galaxies are consistent with having the same $n$ distribution.
Therefore, we suggest that the compact SF1 galaxies undergo a gentle evolution to become PSBs in dense environments, without 
significant structural transformation.

\section{Discussion}
\label{sec:discussion}

 \begin{figure*}
 	\begin{center}
 		\includegraphics[width=0.9\textwidth]{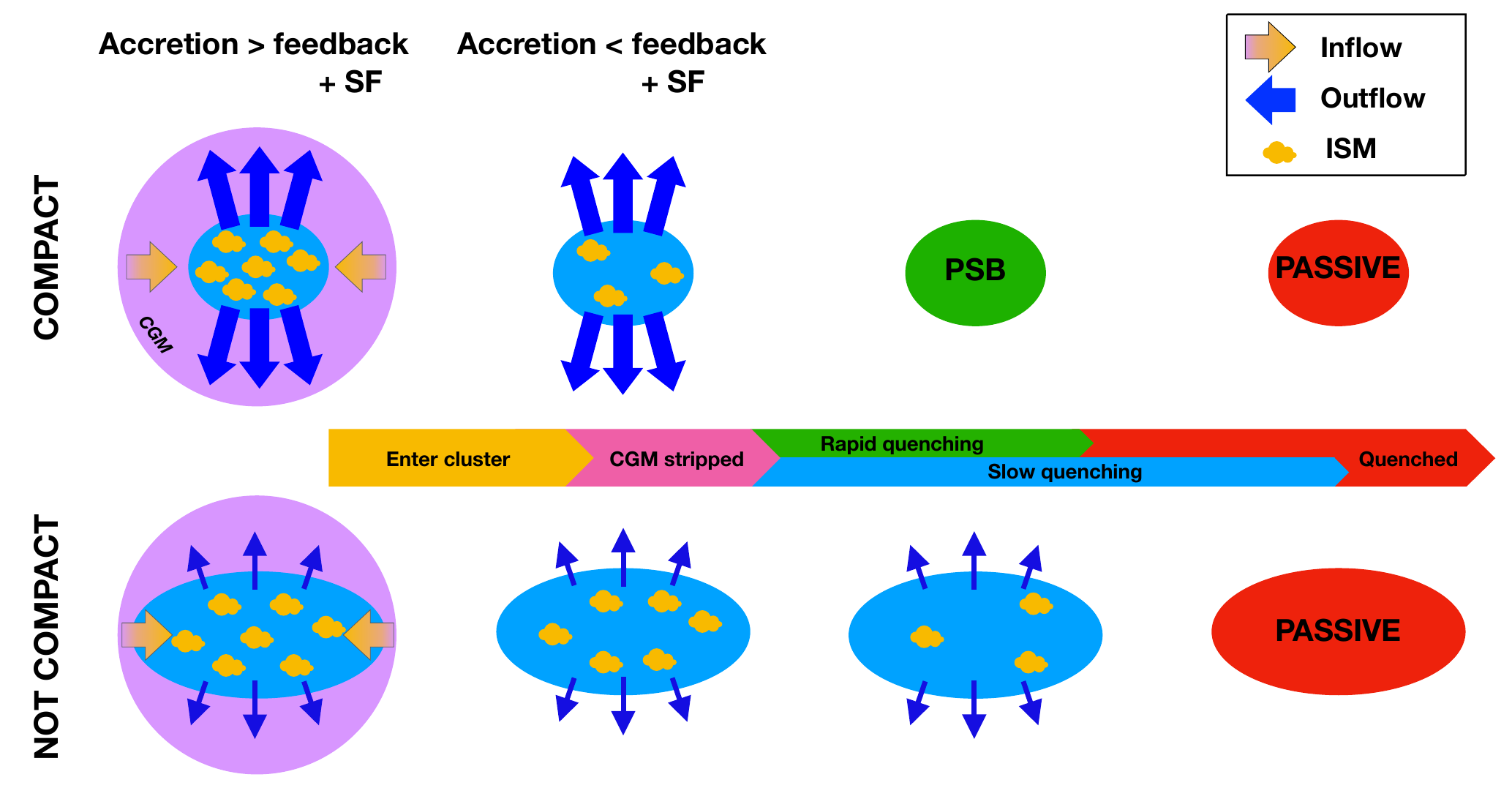}
		\vspace{-0pt}
 		\caption{Diagram illustrating the different environmental quenching pathways followed by high-SSFR galaxies depending on whether they are compact (top row) or not (bottom row). The ellipses represent the location of the stars, the yellow clouds represent the ISM and the purple circles are the CGMs. The blue arrows pointing outwards from the galactic disc represent the outflows, stronger on the top sequence. The yellow arrows pointing inward represent the inflow of gas, which stops immediately as the galaxy comes in contact with the ICM. In the compact SF1 case the strong outflows expel most of the ISM in a short timescale, which rapidly quenches the galaxy leading to the PSB phase. Non-compact SF1 galaxies host weaker outflows, therefore they are able to sustain their star formation over a longer period of time. However, the end state for both pathways is a red, quiescent galaxy.}
 		\label{fig:model}
 	\end{center}
 \end{figure*}

\subsection{The effect of the group environment on the star-forming population}
\label{sec:group SF1}

In Section~\ref{sec:largeSF1}, we showed that the remaining high-SSFR galaxies (SF1 galaxies) in clusters are
on average larger than the SF1 galaxies located in the field.  We also show that this trend is most likely
driven by a lack of SF1 galaxies with small $R_e$ at a given stellar mass, see
Figs.~\ref{fig:subtracted mass-size}~and~\ref{fig:subtracted mass-sersic}.  There are two plausible
explanations for this observation: 1)~environment affects SF1 galaxies in such way that their $R_e$ increases;
or 2)~compact galaxies are being preferentially quenched in the cluster environment.  We expand on these
scenarios below.

Previous observational work has found that elliptical systems tend to be larger in high density environments
than in the field at $z\gtrsim1$ \citep{cooper2012, lani2013}.  From a theoretical point of view, this has
been explained through either the repeated interaction between galaxies (harassment) or dry mergers that take
place in crowded environments \citep{vandokkum2005, shankar2013, oogi2013}.  However, major mergers and
harassment are thought to disrupt galactic discs and lead to an enhancement of the bulge component
\citep{toomre1972, farouki1981, moore1996, gonzalez-garcia2005, aceves2006}.  Consequently, major merging and
harassment do not provide a viable explanation for the large sizes of the cluster SF1 population, which
consists mainly of star-foming discs with typical S\'ersic indices $n\sim1$ (see Fig.~\ref{fig:subtracted mass-sersic}). Conversely, minor galaxy 
mergers are thought to enable the growth of the disc component \citep{younger2007, naab2009, silchenko2011}. 
However, we expect SF1 galaxies to evolve into PSBs through environmental quenching \citep{socolovsky2018}, and PSBs are compact. Therefore, we cannot discard the possibility of two independent processes acting simultaneously on the SF1 population: minor mergers may be responsible for the increase in size of cluster SF1 galaxies and major gas-rich mergers (e.g. between two SF1 galaxies) might be quenching them into compact PSBs \citep{wild2016}. Note that these two processes are disconnected from each other, i.e. galaxies may undergo one of them rather than one after the other. 

The main weakness of the major merger hypothesis is the observed range of S\'ersic indices for both SF1 and PSB in clusters. The median S\'ersic index of cluster PSBs is $n\sim1.5$, which is low for a post-major merger scenario \citep{gonzalez-garcia2005}. Although some simulations have shown that a disc can form after a major wet merger \citep{athanassoula2016}, the time required for this to occur is significantly longer than the expected duration of the PSB phase ($\lesssim1$~Gyr; \citealt{wild2016, socolovsky2018}). At $z<0.1$, PSBs have high S\'ersic index values and are thought to be major merger remnants that tend to reside in low-density environments \citep{zabludoff1996, blake2004, pawlik2018}. In contrast we suggest our PSBs are originated via some kind of gentle gas removal in galaxy clusters. Hence, these results are not contradictory. 
On the other hand, minor mergers with dwarf galaxies that we cannot observe provide a feasible explanation. However, this may be difficult to reconcile with the observed S\'ersic index of SF1 galaxies ($n\sim1$).

Instead, we favour the hypothesis in which compact SF1 galaxies are preferentially quenched 
in dense environments. This naturally leads to the remaining SF1 population appearing on average larger in the cluster while 
the quenched galaxies (PSBs) are smaller than the typical SF1 in the field. 
This preferential quenching of compact objects is hard to 
reconcile with the environmental mechanisms previously mentioned. For example, ram-pressure stripping 
and tidal interactions are expected to act more efficiently in more extended galaxies, 
with shallower gravitational potentials so that the gas is more easily disturbed 
\citep{bothun1993, abadi1999, moore1999}. Scenarios involving quenching induced by galaxy mergers were 
also considered, but these are not expected to depend on galaxy size.

Given that purely environmental processes fail to describe our results, we suggest that the rapid environmental quenching of compact SF1 galaxies is a combination of both, internal and external mechanisms. Our hypothesis, summarized in Fig.~\ref{fig:model}, is based on a ``bathtub''-type model \citep{bouche2010}, in which the star formation in a galaxy is regulated by the balance between gas inflows and outflows. 
Broadly speaking, gas in galaxies is present in two phases: a cold reservoir, and hot reservoir. The cold gas reservoir (or interstellar medium, ISM) corresponds to the dense gas typically found within the disc, and represents the instantaneous fuel for star formation. The hot gas reservoir refers to the extended halo of diffuse gas in which the galaxy is embedded (circumgalactic medium, CGM). The gas in the CGM is too hot to collapse into stars but has the potential to cool down with time and feed the ISM through cold streams, for this reason the CGM is also referred to as the long-term gas reservoir. 

The inflows (represented with yellow arrows in Fig.~\ref{fig:model}) consist of gas from the cosmic web being accreted by the galaxy. On its infall, this gas forms the CGM. As it cools down, it migrates inward.
In contrast, outflows (blue arrows in Fig.~\ref{fig:model}) send gas from the ISM back into the CGM. These outflows could be driven by stellar, supernovae or AGN feedback. Nevertheless, the expelled gas can be recycled after some cooling time, when it is reaccreted into the ISM. 
However, if the galaxy becomes a satellite in a group/cluster, the CGM is largely stripped away via interaction with the intra-cluster medium (\citealt{larson1980}). The ICM also halts the accretion of gas from the cosmic web, so that cluster galaxies are left only with their short-term reservoir to fuel star formation.
Although all galaxies have their hot gas reservoir stripped away almost instantaneously, this has no immediate effect on the ongoing star formation. Therefore, those galaxies with the highest SSFRs and/or strongest galactic outflows will deplete their cold gas reservoir faster and, consequently, quench. This scenario is similar to the `overconsumption' process described in \citet{mcgee2014}, which is proposed to rapidly quench satellite galaxies at $z\sim1.5$.

Some studies have found that compact galaxies are more efficient at transforming gas into stars \citep{young1999}. Similarly, at fixed SFR, compact in size means higher star-formation surface density, which is associated with stronger outflows  \citep{heckman1990}. In this study, we do not find higher SFRs in compact SF1 galaxies but they may host stronger stellar-wind-driven outflows. 
From a theoretical viewpoint, the strength of these super-winds scales with star-formation rate density ($\Sigma_{\rm SFR}$). 
Compact SF1 galaxies have higher $\Sigma_{\rm SFR}$ due to their compact nature and the same SFR as the rest of SF1 galaxies, therefore, they are expected to produce stronger outflows (top row of Fig.~\ref{fig:model}). On the other hand, the rest of the SF1 population (i.e. not compact; second row of Fig.~\ref{fig:model}) may have more modest outflows, which would allow them to stay star-forming over longer timescales before they also run out of fuel (``delayed-then-rapid'' environmental quenching scenario, \citealt{wetzel2013}). This theory provides a successful explanation for why compact SF1 galaxies quench faster than their more extended counterparts. They are more efficient at evacuating their cold gas reservoir after the cluster environment prevents the replenishment of gas by blocking the inflows.

In summary, stellar and supernova winds in combination with the interaction with the ICM may cause the rapid 
quenching predicted by \citet{socolovsky2018} for SF1 galaxies in overdense environments. This hypothesis anticipates 
that environmental quenching does not trigger significant structural evolution. PSBs appear, on 
average, more compact than the general SF1 population because they are primarily the descendants of the compact SF1 galaxies.

This theory predicts stronger outflows in compact SF1 galaxies than in large ones. This can be tested 
by looking at outflow signatures in spectra of galaxies above and below the mass--size relation 
of the field SF1 sample.

\subsection{Spectral analysis: evidence for strong outflows in compact star-forming galaxies}
\label{sec:test}

In this study, we find evidence that compact galaxies with high SSFR (SF1s), are more susceptible to being
quenched in the cluster environment.  We hypothesise that this result could be explained by a combination of
both environmental and secular processes in the following scenario: i)~upon cluster infall, interaction with
the ICM removes the galaxy's extended hot gas reservoir, shutting down cosmic accretion; and ii)~the strong
stellar feedback in these compact galaxies causes significant outflows which rapidly expel any remaining cold
gas from the central regions.  This scenario would naturally lead to the rapid quenching of compact SF1
galaxies in clusters and their subsequent evolution into cluster PSBs.

\begin{figure}
\includegraphics[width=0.49\textwidth]{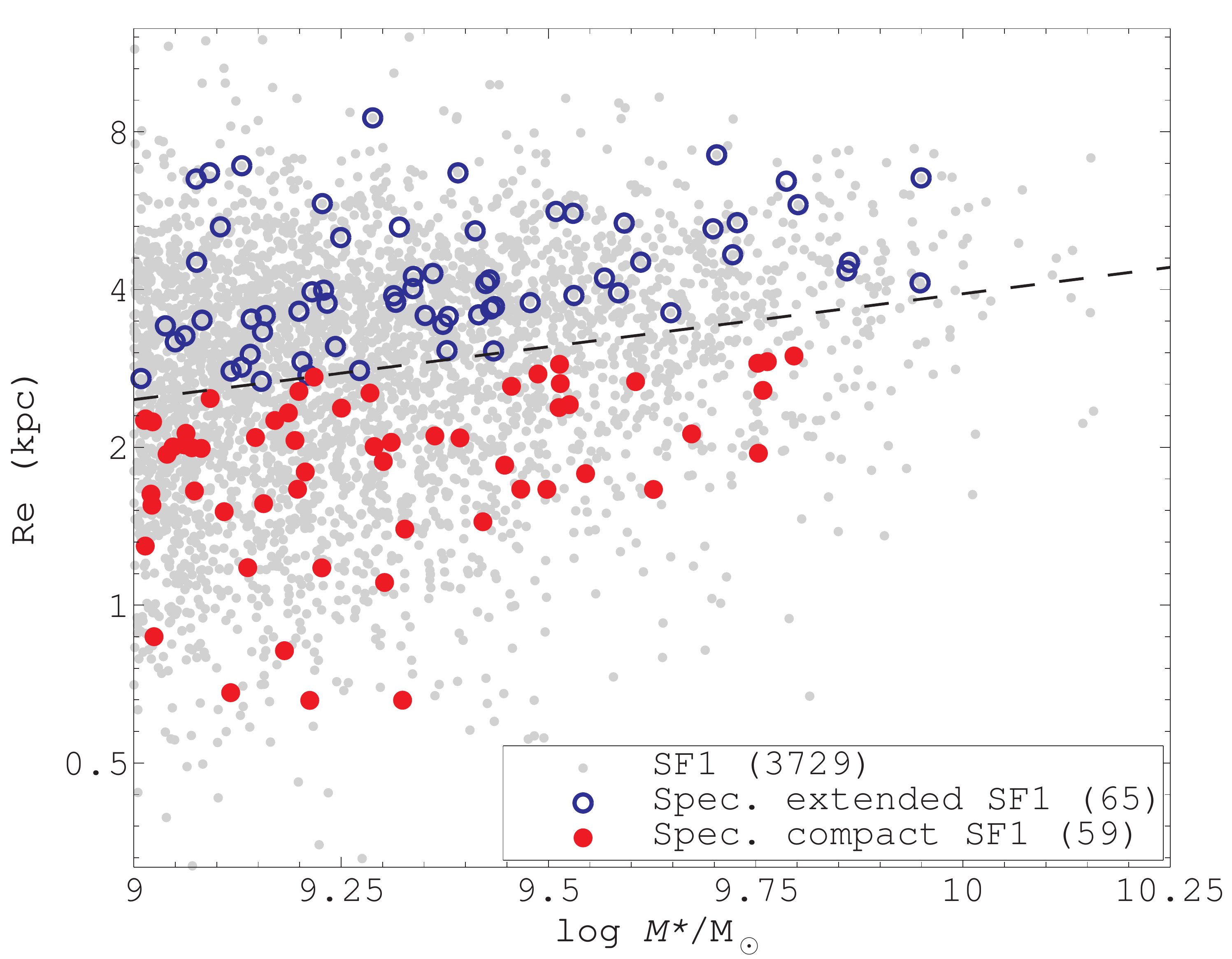}\\
\centering
\caption{\label{spectra-mass-size} The stellar-mass--size relation for SF1 galaxies in the UDS field
($0.5 < z < 1$), showing the sample with available optical spectra from UDSz.  The linear fit to the field
mass--size relation is also shown for reference [black-dashed line; see equation~(\ref{eq:MSRF})].  We define
all galaxies with optical spectra that lie above the fit to the mass--size relation as `extended' (blue
circles), and those that lie below as `compact' (red points).  Relevant sample sizes are shown in the legend.}
\end{figure}

To test this hypothesis we use the available deep optical spectra in the UDS field to determine whether the
strong gaseous outflows required are present in our compact SF1 population.  These spectra are provided by
UDSz, the spectroscopic component of the UDS (ESO Large Programme 180.A-0776, PI: Almaini), which used both
the VIMOS and FORS2 instruments on the ESO VLT to obtain optical spectra for $>3500$ galaxies in the UDS
field \citep[see][]{bradshaw2013, mclure2013}.  For our field/cluster SF1 galaxies, we find that $124$
low-resolution VIMOS spectra ($R\sim200$) are available and that these spectra are evenly distributed
throughout the SF1 mass--size relation (see Fig.~\ref{spectra-mass-size}).  In the following, we define all
galaxies with optical spectra that lie above the fit to the mass--size relation to be `extended', and those
that lie below to be `compact'.  For these spectra, spectroscopic redshifts $z_{\rm spec}$ were obtained
by \cite{bradshaw2013} via {\sc ez} \citep{Garilli2010}, which uses a cross-correlation of spectral
templates.  Optimal solutions were also confirmed using spectral line identification in {\sc sgnaps}
\citep{Paioro2012}.  In this work, we use these redshifts to shift the individual galaxy spectra to their
respective rest-frame (i.e.\ systemic frame).

\begin{figure*}
\includegraphics[width=0.95\textwidth]{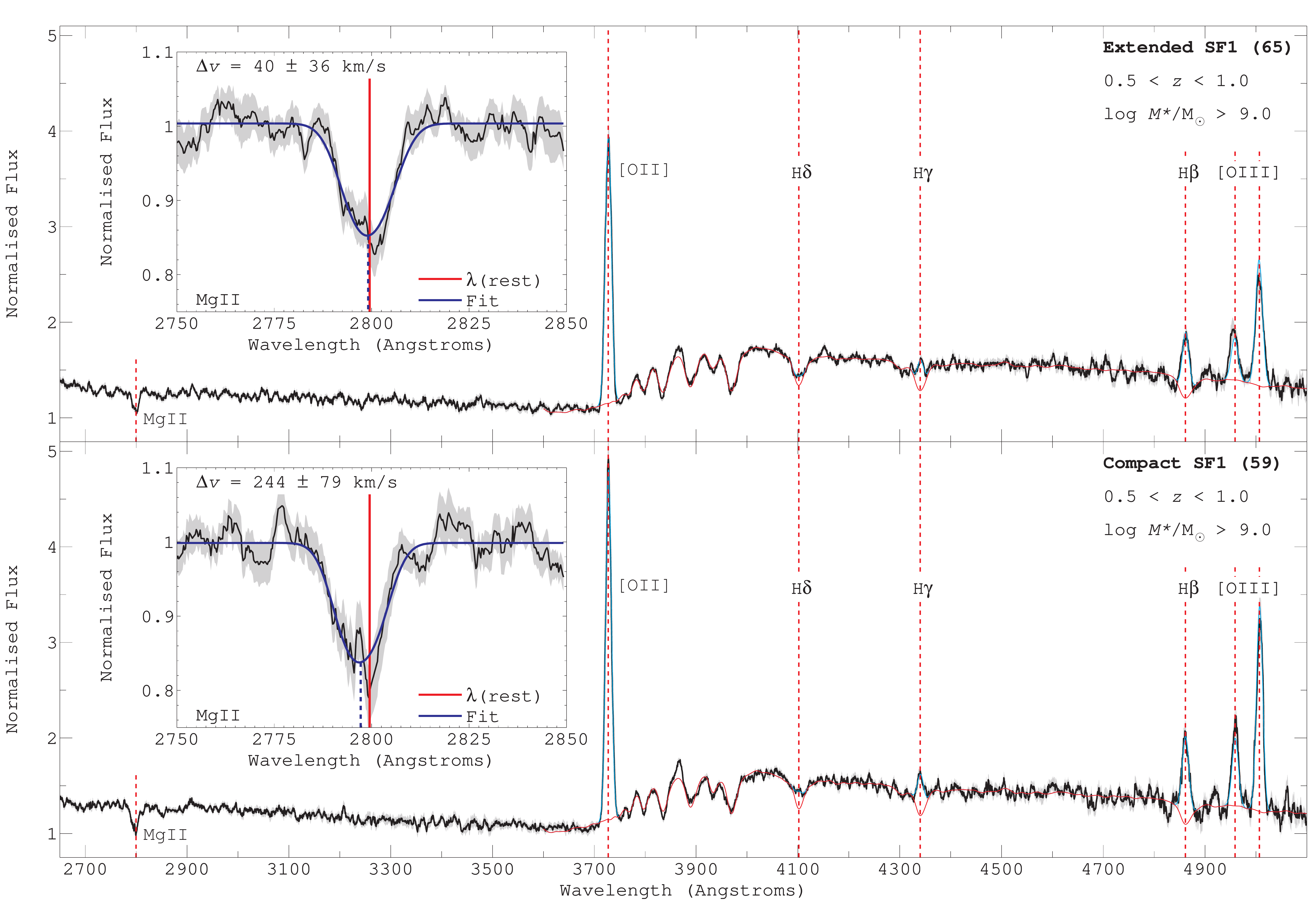}\\
\centering
\caption{\label{spectra}Stacked optical spectra for `extended' and `compact' SF1 galaxies in the UDS field at
$0.5 < z < 1$.  Top panel: a stacked optical spectrum for extended SF1 galaxies (i.e.\ those that lie above
the fit to the mass--size relation in Fig.~\ref{spectra-mass-size}).  For reference, we also show the full
spectral fit obtained from {\sc ppxf} for both the stellar component (red line) and the gas emission lines
(cyan lines).  Bottom panel: an analogous stacked spectrum and {\sc ppxf} fit for compact SF1 galaxies
(i.e.\ those that lie below the fit to the mass--size relation in Fig.~\ref{spectra-mass-size}).  Relevant
sample sizes are shown in the legend.  In each case, the sub-panel shows the best fit to the Mg\,{\sc ii}
absorption profile, using a model comprising a single Gaussian-convolved doublet with a free centroid.  The
central rest-frame wavelength of the Mg\,{\sc ii} doublet ($\lambda 2799.5$\,\AA; red line) and corresponding
offset in the best-fit ($\Delta\lambda$; blue-dashed line) is shown for reference.  These fits yield the
typical velocity offset $\Delta{v}$ of the Mg\,{\sc ii} doublet from the systemic redshift.  For extended SF1
galaxies, we find this velocity offset to be minimal ($\Delta{v} = 40\pm36\rm\,km\,s^{-1}$).  In contrast,
for compact SF1 galaxies, we require a more significant blue-shift in the doublet to explain the structure of
the Mg\,{\sc ii} profile, and therefore a stronger velocity offset ($\Delta{v} = 244\pm79\rm\,km\,s^{-1}$).
These results imply the presence of stronger galactic-scale outflows in more compact SF1 galaxies.}
\end{figure*}

In order to determine the presence of gaseous outflows, we use the Mg\,{\sc ii} absorption doublet
($\lambda\lambda\,2796$, $2803$\,\AA), which is a sensitive tracer of low-ionisation interstellar gas.  The
detection of a blue-shifted component to this absorption feature is generally indicative of galactic-scale
outflows along the line-of-sight to the observer.  Unfortunately, the signal-to-noise (S/N) in the VIMOS
spectra is not sufficient to reliably determine the structure of the Mg\,{\sc ii} profile on an individual
galaxy basis.  We therefore increase the effective S/N via a stacking analysis, combining the individual
rest-frame spectra following an optimised flux normalisation.  For this we generate two median-stacked
spectra: i)~a red-optimised stack ($\lambda > 3500$\,\AA), using a flux normalisation over the Balmer break
region; and ii) a blue-optimised stack ($\lambda < 3700$\,\AA), using a flux normalisation over the
Mg\,{\sc ii} continuum.  For the blue-optimised stack, we also apply an upper $2\sigma$ clip to individual
spectra that deviate from the median flux within the Mg\,{\sc ii} region ($2775 < \lambda < 2825$\,\AA).
This clipping removes a handful of spectra ($<10\%$) that exhibit Mg\,{\sc ii} emission, which would
otherwise bias our stacking analysis.  The final median-stacked spectrum is a splice of the red- and
blue-optimised stacks.  The median-stacked spectra for both our `extended' and `compact' SF1 galaxies are
shown in Fig.~\ref{spectra}.

For both our median stacks, we also perform a full spectral fit (stellar component plus gas emission lines)
using the penalized pixel-fitting method ({\sc ppxf}; \citealt{cappellari2004, cappellari2017}) and the
MILES spectral templates \citep{vazdekis2010}.  Note that due to the range of these spectral templates,
our {\sc ppxf} fits are limited to $\lambda > 3540$\,\AA\ (see Fig.~\ref{spectra}).  We find that our
median-stacks are well-modelled by the resultant spectral fits, which also clearly demonstrate that several
spectral features (e.g.\ Balmer lines, [O\,{\sc ii}], [O\,{\sc iii}]) are all well-centred with respect to
their expected wavelengths.  For Mg\,{\sc ii} this indicates that an observed offset $\Delta\lambda$ from
the systemic-frame wavelength ($\lambda\lambda\,2796$, $2803$\,\AA) is unlikely to be introduced by any
uncertainty in our stacking procedure or individual spectroscopic redshifts.  Consequently, this suggests
that if such offset is observed, it will be related to a genuine velocity offset $\Delta{v}$ in the relevant
absorbing gas from the systemic redshift, and therefore imply the presence of galactic-scale outflows.

In Fig.~\ref{spectra}, an initial comparison of our median stacks reveals a significant difference in the
nature of the Mg\,{\sc ii} absorption between our compact and extended SF1 galaxies.  With respect to the
central systemic-frame wavelength of the Mg\,{\sc ii} doublet ($\lambda 2799.5$\,\AA), for extended SF1s the
absorption profile is well-centred, while for compact SF1s there is a clear offset towards bluer wavelengths.
To determine the significance of this potential offset, we use the following procedure.  We model the
Mg\,{\sc ii} absorption profile using a single component, consisting of a Gaussian-convolved doublet with a
free centroid wavelength.  The doublet itself has a fixed line ratio (i.e.\ 1.1:1; as observed for high-$z$
star-forming galaxies; \citealt{weiner2009}) and an intrinsic narrow width for each line.  In the fitting
process this doublet is convolved with a Gaussian to model the instrument response, which is necessary since
the Mg\,{\sc ii} doublet is unresolved in our low-resolution spectra.  The Gaussian used for convolution is
either fixed to the FWHM of the O\,{\sc ii} emission, or left as a free component in the fit (both cases of
which yield consistent results in this study).  This simple model yields the offset $\Delta\lambda$ of the
Mg\,{\sc ii} doublet with respect to the systemic-frame wavelength, which can be used to determine a
characteristic velocity offset $\Delta{v}$ from the systemic redshift.  For our stacked spectra, this
velocity offset represents an estimate of the typical outflow velocity in the low-ionisation gas for our
galaxy populations.  The $1\sigma$ uncertainties in these velocity measurements are determined using the
variance between analogous fits performed on $1000$ simulated spectra generated via a bootstrap analysis.

For our `extended' and `compact' SF1 galaxies, the relevant fits to the Mg\,{\sc ii} profile are presented
in Fig.~\ref{spectra}.  In each case, we determine the velocity offset $\Delta{v}$ of the Mg\,{\sc ii}
absorption profile with respect to the systemic redshift.  In the case of extended SF1 galaxies, we find this
velocity offset to be minimal ($\Delta{v} = 40\pm36\rm\,km\,s^{-1}$).  This indicates that no significant
outflowing (i.e.\ blue-shifted) component of low-ionisation gas is present in these galaxies.  However, for
compact SF1 galaxies, we find a significant excess of blue-shifted absorption indicative of gaseous outflows.
In this case, our best-fit model yields a significant velocity offset in the Mg\,{\sc ii} absorption profile
($\Delta{v} = 244\pm79\rm\,km\,s^{-1}$).  This indicates that the strong stellar feedback inherent to these
compact star-forming galaxies is likely causing strong galactic-scale outflows or winds in their interstellar
medium.  Finally, we note that more complex models involving two components (e.g.\ one fixed at the
rest-frame wavelength for the systemic absorption, and another with a free centroid to model the outflow;
see e.g.\ \citealt{bradshaw2013}) were also explored, and that these all yield consistent results, to within
the fit uncertainties.  We also note that in each case, the use of either median- or mean-stacked spectra in
our analysis leads to consistent Mg\,{\sc ii} profiles and $\Delta{v}$ measurements, and has no significant
impact on the results of this study.

Taken together, these results indicate that strong galactic-scale outflows are commonplace in compact SF1
galaxies, but not a significant factor in the more extended SF1 galaxies. This supports our hypothesis that
when SF1 galaxies infall to the cluster environment and have their extended gas reservoirs removed by ICM
interactions, the subsequent evolution is strongly dependent on the compactness of the galaxy.  For extended
galaxies, the lack of strong outflows leads to the galaxy retaining its cold gas disc and therefore the
continuation of star formation.  In contrast, for compact galaxies the stronger outflows present will quickly
lead to the removal of the remaining cold gas disc, which would result in the rapid quenching of star
formation and the subsequent evolution of these galaxies into cluster PSBs.

\section{Conclusions}
\label{sec:conclusions}

We present the first evidence that the structure of galaxies with high SSFRs (SF1s) differs with environment
at $0.5 < z < 1.0$. Using $K$-band structural parameters available for the UDS, we find that high-SSFR
galaxies in clusters are typically larger than analogous galaxies in the field.  In recent work, we found
that these galaxies are strongly depleted in dense environments and undergo rapid quenching to become cluster
PSBs \citep[see][]{socolovsky2018}. We therefore suggest that the observed difference in size is caused by
the preferential quenching of compact galaxies in dense environments. We summarise our main findings as
follows:

\begin{enumerate}

\item Using the mass--size relation, we find that galaxies with high SSFR in the cluster environment are on
average larger than their counterparts in the field.

\vspace{0.1cm}

\item Examining the distribution of effective radii $R_{\rm e}$, we find that the difference in size is
likely to be driven by a lack of compact SF1 galaxies in clusters.  This suggests a preferential
environmental quenching of the most compact galaxies.  From a similar analysis of the distribution in
S\'ersic indices, we infer that the missing compact SF1 galaxies had higher S\'ersic index, $n$, than the
typical SF1 galaxy in the field.

\vspace{0.1cm}

\item We find that the structural parameters of the missing compact SF1 galaxies are compatible with those of
the cluster PSB population.  Building on the work of \citet{socolovsky2018}, this suggests that compact SF1s
are the main progenitors of cluster PSBs, rather than the SF1 population as a whole.  These galaxies are
rapidly quenched and evolve into the PSB population with no significant structural evolution.

\end{enumerate}

Taken together, these results indicate that rapid quenching within clusters is
size-dependent at $0.5 < z < 1.0$. This may explain why cluster PSBs are significantly smaller than the typical SF1 galaxy or
indeed the general star forming population \citep{maltby2018}.

Regarding the quenching mechanisms, we suggest that the most likely scenario combines secular and
environmental processes.  The interaction with the ICM blocks the inflow of gas into the galaxy, which
results in the exhaustion of the gas reservoir through star formation and outflows.  Therefore, compact SF1
galaxies, which have higher surface star formation densities (similar SFR in a smaller radius), rapidly run
out of fuel due to their stronger outflows.  This hypothesis is supported by the spectroscopic data available
for $124$ of the SF1 galaxies.  The spectra show evidence that the Mg\,{\sc ii} absorption feature contains a
significant blue-shifted component, indicative of outflows, in those galaxies that lie below the field SF1
mass--size relation in comparison to those above it. This provides evidence supporting our model, suggesting
that compact SF1 galaxies tend to host stronger galactic outflows. 

In conclusion, we find evidence for size-dependent environmental quenching in clusters at $0.5 < z < 1.0$.
Our results show that compact star-forming galaxies are preferentially and rapidly quenched in clusters to
become PSBs.

\section{Acknowledgements}

This work uses data from ESO telescopes at the Paranal Observatory
(programme 180.A-0776; PI: Almaini). We are grateful
to the staff at UKIRT for their tireless efforts in ensuring the
success of the UDS project. We were most
fortunate to have the opportunity to conduct observations from this
mountain. MS acknowledges support from IAC and STFC. VW acknowledges 
support from the European Research Council Starting grant (SEDmorph, P.I. V. Wild).

\bibliographystyle{mnras} 
\bibliography{biblio}

\begin{thebibliography}{}
\makeatletter
\relax
\def\mn@urlcharsother{\let\do\@makeother \do\$\do\&\do\#\do\^\do\_\do\%\do\~}
\def\mn@doi{\begingroup\mn@urlcharsother \@ifnextchar [ {\mn@doi@}
  {\mn@doi@[]}}
\def\mn@doi@[#1]#2{\def\@tempa{#1}\ifx\@tempa\@empty \href
  {http://dx.doi.org/#2} {doi:#2}\else \href {http://dx.doi.org/#2} {#1}\fi
  \endgroup}
\def\mn@eprint#1#2{\mn@eprint@#1:#2::\@nil}
\def\mn@eprint@arXiv#1{\href {http://arxiv.org/abs/#1} {{\tt arXiv:#1}}}
\def\mn@eprint@dblp#1{\href {http://dblp.uni-trier.de/rec/bibtex/#1.xml}
  {dblp:#1}}
\def\mn@eprint@#1:#2:#3:#4\@nil{\def\@tempa {#1}\def\@tempb {#2}\def\@tempc
  {#3}\ifx \@tempc \@empty \let \@tempc \@tempb \let \@tempb \@tempa \fi \ifx
  \@tempb \@empty \def\@tempb {arXiv}\fi \@ifundefined
  {mn@eprint@\@tempb}{\@tempb:\@tempc}{\expandafter \expandafter \csname
  mn@eprint@\@tempb\endcsname \expandafter{\@tempc}}}

\bibitem[\protect\citeauthoryear{{Abadi}, {Moore}  \& {Bower}}{{Abadi}
  et~al.}{1999}]{abadi1999}
{Abadi} M.~G.,  {Moore} B.,   {Bower} R.~G.,  1999, \mn@doi [\mnras]
  {10.1046/j.1365-8711.1999.02715.x}, 308, 947

\bibitem[\protect\citeauthoryear{{Aceves}, {Vel{\'a}zquez}  \& {Cruz}}{{Aceves}
  et~al.}{2006}]{aceves2006}
{Aceves} H.,  {Vel{\'a}zquez} H.,   {Cruz} F.,  2006, \mn@doi [\mnras]
  {10.1111/j.1365-2966.2006.11029.x}, 373, 632

\bibitem[\protect\citeauthoryear{{Almaini} et~al.,}{{Almaini}
  et~al.}{2017}]{almaini2017}
{Almaini} O.,  et~al., 2017, \mn@doi [\mnras] {10.1093/mnras/stx1957}, 472,
  1401

\bibitem[\protect\citeauthoryear{{Athanassoula}, {Rodionov}, {Peschken}  \&
  {Lambert}}{{Athanassoula} et~al.}{2016}]{athanassoula2016}
{Athanassoula} E.,  {Rodionov} S.~A.,  {Peschken} N.,   {Lambert} J.~C.,  2016,
  \mn@doi [\apj] {10.3847/0004-637X/821/2/90}, 821, 90

\bibitem[\protect\citeauthoryear{{Baldry}, {Balogh}, {Bower}, {Glazebrook},
  {Nichol}, {Bamford}  \& {Budavari}}{{Baldry} et~al.}{2006}]{baldry2006}
{Baldry} I.~K.,  {Balogh} M.~L.,  {Bower} R.~G.,  {Glazebrook} K.,  {Nichol}
  R.~C.,  {Bamford} S.~P.,   {Budavari} T.,  2006, \mn@doi [\mnras]
  {10.1111/j.1365-2966.2006.11081.x}, 373, 469

\bibitem[\protect\citeauthoryear{{Baldry} et~al.,}{{Baldry}
  et~al.}{2012}]{baldry2012}
{Baldry} I.~K.,  et~al., 2012, \mn@doi [\mnras]
  {10.1111/j.1365-2966.2012.20340.x}, 421, 621

\bibitem[\protect\citeauthoryear{{Balogh}, {Morris}, {Yee}, {Carlberg}  \&
  {Ellingson}}{{Balogh} et~al.}{1997}]{balogh1997}
{Balogh} M.~L.,  {Morris} S.~L.,  {Yee} H.~K.~C.,  {Carlberg} R.~G.,
  {Ellingson} E.,  1997, \mn@doi [\apjl] {10.1086/310927}, 488, L75

\bibitem[\protect\citeauthoryear{{Bamford} et~al.,}{{Bamford}
  et~al.}{2009}]{bamford2009}
{Bamford} S.~P.,  et~al., 2009, \mn@doi [\mnras]
  {10.1111/j.1365-2966.2008.14252.x}, 393, 1324

\bibitem[\protect\citeauthoryear{{Barden}, {H{\"a}u{\ss}ler}, {Peng},
  {McIntosh}  \& {Guo}}{{Barden} et~al.}{2012}]{barden2012}
{Barden} M.,  {H{\"a}u{\ss}ler} B.,  {Peng} C.~Y.,  {McIntosh} D.~H.,   {Guo}
  Y.,  2012, \mn@doi [\mnras] {10.1111/j.1365-2966.2012.20619.x}, 422, 449

\bibitem[\protect\citeauthoryear{{Blake} et~al.,}{{Blake}
  et~al.}{2004}]{blake2004}
{Blake} C.,  et~al., 2004, \mn@doi [\mnras] {10.1111/j.1365-2966.2004.08351.x},
  355, 713

\bibitem[\protect\citeauthoryear{{Bothun}, {Schombert}, {Impey}, {Sprayberry}
  \& {McGaugh}}{{Bothun} et~al.}{1993}]{bothun1993}
{Bothun} G.~D.,  {Schombert} J.~M.,  {Impey} C.~D.,  {Sprayberry} D.,
  {McGaugh} S.~S.,  1993, \mn@doi [\aj] {10.1086/116659}, 106, 530

\bibitem[\protect\citeauthoryear{{Bouch{\'e}} et~al.,}{{Bouch{\'e}}
  et~al.}{2010}]{bouche2010}
{Bouch{\'e}} N.,  et~al., 2010, \mn@doi [\apj] {10.1088/0004-637X/718/2/1001},
  718, 1001

\bibitem[\protect\citeauthoryear{{Bradshaw} et~al.,}{{Bradshaw}
  et~al.}{2013}]{bradshaw2013}
{Bradshaw} E.~J.,  et~al., 2013, \mn@doi [\mnras] {10.1093/mnras/stt715}, 433,
  194

\bibitem[\protect\citeauthoryear{{Brammer}, {van Dokkum}  \& {Coppi}}{{Brammer}
  et~al.}{2008}]{brammer2008}
{Brammer} G.~B.,  {van Dokkum} P.~G.,   {Coppi} P.,  2008, \mn@doi [\apj]
  {10.1086/591786}, 686, 1503

\bibitem[\protect\citeauthoryear{{Bruzual} \& {Charlot}}{{Bruzual} \&
  {Charlot}}{2003}]{bruzual2003}
{Bruzual} G.,  {Charlot} S.,  2003, \mn@doi [\mnras]
  {10.1046/j.1365-8711.2003.06897.x}, 344, 1000

\bibitem[\protect\citeauthoryear{{Cappellari}}{{Cappellari}}{2017}]{cappellari%
2017}
{Cappellari} M.,  2017, \mn@doi [\mnras] {10.1093/mnras/stw3020}, 466, 798

\bibitem[\protect\citeauthoryear{{Cappellari} \& {Emsellem}}{{Cappellari} \&
  {Emsellem}}{2004}]{cappellari2004}
{Cappellari} M.,  {Emsellem} E.,  2004, \mn@doi [\pasp] {10.1086/381875}, 116,
  138

\bibitem[\protect\citeauthoryear{{Cebri{\'a}n} \& {Trujillo}}{{Cebri{\'a}n} \&
  {Trujillo}}{2014}]{cebrian2014}
{Cebri{\'a}n} M.,  {Trujillo} I.,  2014, \mn@doi [\mnras]
  {10.1093/mnras/stu1375}, 444, 682

\bibitem[\protect\citeauthoryear{{Chabrier}}{{Chabrier}}{2003}]{chabrier2003}
{Chabrier} G.,  2003, \mn@doi [\pasp] {10.1086/376392}, 115, 763

\bibitem[\protect\citeauthoryear{{Cooper} et~al.,}{{Cooper}
  et~al.}{2012}]{cooper2012}
{Cooper} M.~C.,  et~al., 2012, \mn@doi [\mnras]
  {10.1111/j.1365-2966.2011.19938.x}, 419, 3018

\bibitem[\protect\citeauthoryear{{Drory} et~al.,}{{Drory}
  et~al.}{2009}]{drory2009}
{Drory} N.,  et~al., 2009, \mn@doi [\apj] {10.1088/0004-637X/707/2/1595}, 707,
  1595

\bibitem[\protect\citeauthoryear{{Faber}}{{Faber}}{1973}]{faber1973}
{Faber} S.~M.,  1973, \mn@doi [\apj] {10.1086/151881}, 179, 423

\bibitem[\protect\citeauthoryear{{Farouki} \& {Shapiro}}{{Farouki} \&
  {Shapiro}}{1981}]{farouki1981}
{Farouki} R.,  {Shapiro} S.~L.,  1981, \mn@doi [\apj] {10.1086/158563}, 243, 32

\bibitem[\protect\citeauthoryear{{Furusawa} et~al.,}{{Furusawa}
  et~al.}{2008}]{furusawa2008}
{Furusawa} H.,  et~al., 2008, \mn@doi [\apjs] {10.1086/527321}, 176, 1

\bibitem[\protect\citeauthoryear{{Garilli}, {Fumana}, {Franzetti}, {Paioro},
  {Scodeggio}, {Le F{\`e}vre}, {Paltani}  \& {Scaramella}}{{Garilli}
  et~al.}{2010}]{Garilli2010}
{Garilli} B.,  {Fumana} M.,  {Franzetti} P.,  {Paioro} L.,  {Scodeggio} M.,
  {Le F{\`e}vre} O.,  {Paltani} S.,   {Scaramella} R.,  2010, \mn@doi [\pasp]
  {10.1086/654903}, 122, 827

\bibitem[\protect\citeauthoryear{{Gonz{\'a}lez-Garc{\'{\i}}a} \&
  {Balcells}}{{Gonz{\'a}lez-Garc{\'{\i}}a} \&
  {Balcells}}{2005}]{gonzalez-garcia2005}
{Gonz{\'a}lez-Garc{\'{\i}}a} A.~C.,  {Balcells} M.,  2005, \mn@doi [\mnras]
  {10.1111/j.1365-2966.2005.08706.x}, 357, 753

\bibitem[\protect\citeauthoryear{{Gunn} \& {Gott}}{{Gunn} \&
  {Gott}}{1972}]{gunn1972}
{Gunn} J.~E.,  {Gott} III J.~R.,  1972, \mn@doi [\apj] {10.1086/151605}, 176, 1

\bibitem[\protect\citeauthoryear{{Hartley} et~al.,}{{Hartley}
  et~al.}{2013}]{hartley2013}
{Hartley} W.~G.,  et~al., 2013, \mn@doi [\mnras] {10.1093/mnras/stt383}, 431,
  3045

\bibitem[\protect\citeauthoryear{{Heckman}, {Armus}  \& {Miley}}{{Heckman}
  et~al.}{1990}]{heckman1990}
{Heckman} T.~M.,  {Armus} L.,   {Miley} G.~K.,  1990, \mn@doi [\apjs]
  {10.1086/191522}, 74, 833

\bibitem[\protect\citeauthoryear{{Kelkar}, {Arag{\'o}n-Salamanca}, {Gray},
  {Maltby}, {Vulcani}, {De Lucia}, {Poggianti}  \& {Zaritsky}}{{Kelkar}
  et~al.}{2015}]{kelkar2015}
{Kelkar} K.,  {Arag{\'o}n-Salamanca} A.,  {Gray} M.~E.,  {Maltby} D.,
  {Vulcani} B.,  {De Lucia} G.,  {Poggianti} B.~M.,   {Zaritsky} D.,  2015,
  \mn@doi [\mnras] {10.1093/mnras/stv670}, 450, 1246

\bibitem[\protect\citeauthoryear{{Kelvin} et~al.,}{{Kelvin}
  et~al.}{2012}]{kelvin2012}
{Kelvin} L.~S.,  et~al., 2012, \mn@doi [\mnras]
  {10.1111/j.1365-2966.2012.20355.x}, 421, 1007

\bibitem[\protect\citeauthoryear{{Lani} et~al.,}{{Lani}
  et~al.}{2013}]{lani2013}
{Lani} C.,  et~al., 2013, \mn@doi [\mnras] {10.1093/mnras/stt1275}, 435, 207

\bibitem[\protect\citeauthoryear{{Larson}, {Tinsley}  \& {Caldwell}}{{Larson}
  et~al.}{1980}]{larson1980}
{Larson} R.~B.,  {Tinsley} B.~M.,   {Caldwell} C.~N.,  1980, \mn@doi [\apj]
  {10.1086/157917}, 237, 692

\bibitem[\protect\citeauthoryear{{Maltby} et~al.,}{{Maltby}
  et~al.}{2010}]{maltby2010}
{Maltby} D.~T.,  et~al., 2010, \mn@doi [\mnras]
  {10.1111/j.1365-2966.2009.15953.x}, 402, 282

\bibitem[\protect\citeauthoryear{{Maltby} et~al.,}{{Maltby}
  et~al.}{2016}]{maltby2016}
{Maltby} D.~T.,  et~al., 2016, \mn@doi [\mnras] {10.1093/mnrasl/slw057}, 459,
  L114

\bibitem[\protect\citeauthoryear{{Maltby}, {Almaini}, {Wild}, {Hatch},
  {Hartley}, {Simpson}, {Rowlands}  \& {Socolovsky}}{{Maltby}
  et~al.}{2018}]{maltby2018}
{Maltby} D.~T.,  {Almaini} O.,  {Wild} V.,  {Hatch} N.~A.,  {Hartley} W.~G.,
  {Simpson} C.,  {Rowlands} K.,   {Socolovsky} M.,  2018, \mn@doi [\mnras]
  {10.1093/mnras/sty1794}, 480, 381

\bibitem[\protect\citeauthoryear{{McGee}, {Bower}  \& {Balogh}}{{McGee}
  et~al.}{2014}]{mcgee2014}
{McGee} S.~L.,  {Bower} R.~G.,   {Balogh} M.~L.,  2014, \mn@doi [\mnras]
  {10.1093/mnrasl/slu066}, 442, L105

\bibitem[\protect\citeauthoryear{{McLure} et~al.}{{McLure}
  et~al.}{2013}]{mclure2013}
{McLure} R.~J.,  et~al., 2013, \mn@doi [\mnras] {10.1093/mnras/sts092}, 428,
  1088

\bibitem[\protect\citeauthoryear{{Moore}, {Katz}, {Lake}, {Dressler}  \&
  {Oemler}}{{Moore} et~al.}{1996}]{moore1996}
{Moore} B.,  {Katz} N.,  {Lake} G.,  {Dressler} A.,   {Oemler} A.,  1996,
  \mn@doi [\nat] {10.1038/379613a0}, 379, 613

\bibitem[\protect\citeauthoryear{{Moore}, {Lake}, {Quinn}  \& {Stadel}}{{Moore}
  et~al.}{1999}]{moore1999}
{Moore} B.,  {Lake} G.,  {Quinn} T.,   {Stadel} J.,  1999, \mn@doi [\mnras]
  {10.1046/j.1365-8711.1999.02345.x}, 304, 465

\bibitem[\protect\citeauthoryear{{Moutard} et~al.,}{{Moutard}
  et~al.}{2016}]{moutard2016b}
{Moutard} T.,  et~al., 2016, \mn@doi [\aap] {10.1051/0004-6361/201527294}, 590,
  A103

\bibitem[\protect\citeauthoryear{{Moutard}, {Sawicki}, {Arnouts}, {Golob},
  {Malavasi}, {Adami}, {Coupon}  \& {Ilbert}}{{Moutard}
  et~al.}{2018}]{moutard2018}
{Moutard} T.,  {Sawicki} M.,  {Arnouts} S.,  {Golob} A.,  {Malavasi} N.,
  {Adami} C.,  {Coupon} J.,   {Ilbert} O.,  2018, \mn@doi [\mnras]
  {10.1093/mnras/sty1543}, 479, 2147

\bibitem[\protect\citeauthoryear{{Muzzin} et~al.,}{{Muzzin}
  et~al.}{2013}]{muzzin2013}
{Muzzin} A.,  et~al., 2013, \mn@doi [\apj] {10.1088/0004-637X/777/1/18}, 777,
  18

\bibitem[\protect\citeauthoryear{{Naab}, {Johansson}  \& {Ostriker}}{{Naab}
  et~al.}{2009}]{naab2009}
{Naab} T.,  {Johansson} P.~H.,   {Ostriker} J.~P.,  2009, \mn@doi [\apjl]
  {10.1088/0004-637X/699/2/L178}, 699, L178

\bibitem[\protect\citeauthoryear{{Oogi} \& {Habe}}{{Oogi} \&
  {Habe}}{2013}]{oogi2013}
{Oogi} T.,  {Habe} A.,  2013, \mn@doi [\mnras] {10.1093/mnras/sts047}, 428, 641

\bibitem[\protect\citeauthoryear{{Paioro} \& {Franzetti}}{{Paioro} \&
  {Franzetti}}{2012}]{Paioro2012}
{Paioro} L.,  {Franzetti} P.,  2012, {SGNAPS: Software for Graphical
  Navigation, Analysis and Plotting of Spectra}, Astrophysics Source Code
  Library (\mn@eprint {ascl} {1210.005})

\bibitem[\protect\citeauthoryear{{Papovich} et~al.,}{{Papovich}
  et~al.}{2012}]{papovich2012}
{Papovich} C.,  et~al., 2012, \mn@doi [\apj] {10.1088/0004-637X/750/2/93}, 750,
  93

\bibitem[\protect\citeauthoryear{{Pawlik} et~al.,}{{Pawlik}
  et~al.}{2018}]{pawlik2018}
{Pawlik} M.~M.,  et~al., 2018, \mn@doi [\mnras] {10.1093/mnras/sty589}, 477,
  1708

\bibitem[\protect\citeauthoryear{{Peng}, {Ho}, {Impey}  \& {Rix}}{{Peng}
  et~al.}{2002}]{peng2002}
{Peng} C.~Y.,  {Ho} L.~C.,  {Impey} C.~D.,   {Rix} H.-W.,  2002, \mn@doi [\aj]
  {10.1086/340952}, 124, 266

\bibitem[\protect\citeauthoryear{{Peng} et~al.,}{{Peng}
  et~al.}{2010}]{peng2010}
{Peng} Y.-j.,  et~al., 2010, \mn@doi [\apj] {10.1088/0004-637X/721/1/193}, 721,
  193

\bibitem[\protect\citeauthoryear{{Pozzetti} et~al.,}{{Pozzetti}
  et~al.}{2010}]{pozzetti2010}
{Pozzetti} L.,  et~al., 2010, \mn@doi [\aap] {10.1051/0004-6361/200913020},
  523, A13

\bibitem[\protect\citeauthoryear{{S\'ersic}}{{S\'ersic}}{1968}]{sersic1968}
{S\'ersic} J.~L.,  1968, {Atlas de Galaxias Australes}

\bibitem[\protect\citeauthoryear{{Shankar}, {Marulli}, {Bernardi}, {Mei},
  {Meert}  \& {Vikram}}{{Shankar} et~al.}{2013}]{shankar2013}
{Shankar} F.,  {Marulli} F.,  {Bernardi} M.,  {Mei} S.,  {Meert} A.,   {Vikram}
  V.,  2013, \mn@doi [\mnras] {10.1093/mnras/sts001}, 428, 109

\bibitem[\protect\citeauthoryear{{Shen}, {Mo}, {White}, {Blanton}, {Kauffmann},
  {Voges}, {Brinkmann}  \& {Csabai}}{{Shen} et~al.}{2003}]{shen2003}
{Shen} S.,  {Mo} H.~J.,  {White} S.~D.~M.,  {Blanton} M.~R.,  {Kauffmann} G.,
  {Voges} W.,  {Brinkmann} J.,   {Csabai} I.,  2003, \mn@doi [\mnras]
  {10.1046/j.1365-8711.2003.06740.x}, 343, 978

\bibitem[\protect\citeauthoryear{{Sil'Chenko}, {Chilingarian}, {Sotnikova}  \&
  {Afanasiev}}{{Sil'Chenko} et~al.}{2011}]{silchenko2011}
{Sil'Chenko} O.~K.,  {Chilingarian} I.~V.,  {Sotnikova} N.~Y.,   {Afanasiev}
  V.~L.,  2011, \mn@doi [\mnras] {10.1111/j.1365-2966.2011.18665.x}, 414, 3645

\bibitem[\protect\citeauthoryear{{Simpson} et~al.,}{{Simpson}
  et~al.}{2012}]{simpson2012}
{Simpson} C.,  et~al., 2012, \mn@doi [\mnras]
  {10.1111/j.1365-2966.2012.20529.x}, 421, 3060

\bibitem[\protect\citeauthoryear{{Simpson}, {Westoby}, {Arumugam}, {Ivison},
  {Hartley}  \& {Almaini}}{{Simpson} et~al.}{2013}]{simpson2013}
{Simpson} C.,  {Westoby} P.,  {Arumugam} V.,  {Ivison} R.,  {Hartley} W.,
  {Almaini} O.,  2013, \mn@doi [\mnras] {10.1093/mnras/stt940}, 433, 2647

\bibitem[\protect\citeauthoryear{{Socolovsky}, {Almaini}, {Hatch}, {Wild},
  {Maltby}, {Hartley}  \& {Simpson}}{{Socolovsky}
  et~al.}{2018}]{socolovsky2018}
{Socolovsky} M.,  {Almaini} O.,  {Hatch} N.~A.,  {Wild} V.,  {Maltby} D.~T.,
  {Hartley} W.~G.,   {Simpson} C.,  2018, \mn@doi [\mnras]
  {10.1093/mnras/sty312}, 476, 1242

\bibitem[\protect\citeauthoryear{{Tomczak} et~al.,}{{Tomczak}
  et~al.}{2014}]{tomczak2014}
{Tomczak} A.~R.,  et~al., 2014, \mn@doi [\apj] {10.1088/0004-637X/783/2/85},
  783, 85

\bibitem[\protect\citeauthoryear{{Toomre} \& {Toomre}}{{Toomre} \&
  {Toomre}}{1972}]{toomre1972}
{Toomre} A.,  {Toomre} J.,  1972, \mn@doi [\apj] {10.1086/151823}, 178, 623

\bibitem[\protect\citeauthoryear{{Ueda} et~al.,}{{Ueda}
  et~al.}{2008}]{ueda2008}
{Ueda} Y.,  et~al., 2008, \mn@doi [\apjs] {10.1086/591083}, 179, 124

\bibitem[\protect\citeauthoryear{{Vazdekis}, {S{\'a}nchez-Bl{\'a}zquez},
  {Falc{\'o}n-Barroso}, {Cenarro}, {Beasley}, {Cardiel}, {Gorgas}  \&
  {Peletier}}{{Vazdekis} et~al.}{2010}]{vazdekis2010}
{Vazdekis} A.,  {S{\'a}nchez-Bl{\'a}zquez} P.,  {Falc{\'o}n-Barroso} J.,
  {Cenarro} A.~J.,  {Beasley} M.~A.,  {Cardiel} N.,  {Gorgas} J.,   {Peletier}
  R.~F.,  2010, \mn@doi [\mnras] {10.1111/j.1365-2966.2010.16407.x}, 404, 1639

\bibitem[\protect\citeauthoryear{{Weiner} et~al.,}{{Weiner}
  et~al.}{2009}]{weiner2009}
{Weiner} B.~J.,  et~al., 2009, \mn@doi [\apj] {10.1088/0004-637X/692/1/187},
  692, 187

\bibitem[\protect\citeauthoryear{{Wetzel}, {Tinker}, {Conroy}  \& {van den
  Bosch}}{{Wetzel} et~al.}{2013}]{wetzel2013}
{Wetzel} A.~R.,  {Tinker} J.~L.,  {Conroy} C.,   {van den Bosch} F.~C.,  2013,
  \mn@doi [\mnras] {10.1093/mnras/stt469}, 432, 336

\bibitem[\protect\citeauthoryear{{Wild} et~al.,}{{Wild}
  et~al.}{2014}]{wild2014}
{Wild} V.,  et~al., 2014, \mn@doi [\mnras] {10.1093/mnras/stu212}, 440, 1880

\bibitem[\protect\citeauthoryear{{Wild}, {Almaini}, {Dunlop}, {Simpson},
  {Rowlands}, {Bowler}, {Maltby}  \& {McLure}}{{Wild} et~al.}{2016}]{wild2016}
{Wild} V.,  {Almaini} O.,  {Dunlop} J.,  {Simpson} C.,  {Rowlands} K.,
  {Bowler} R.,  {Maltby} D.,   {McLure} R.,  2016, \mn@doi [\mnras]
  {10.1093/mnras/stw1996}, 463, 832

\bibitem[\protect\citeauthoryear{{Yang}, {Mo}  \& {van den Bosch}}{{Yang}
  et~al.}{2009}]{yang2009}
{Yang} X.,  {Mo} H.~J.,   {van den Bosch} F.~C.,  2009, \mn@doi [\apj]
  {10.1088/0004-637X/695/2/900}, 695, 900

\bibitem[\protect\citeauthoryear{{Young}}{{Young}}{1999}]{young1999}
{Young} J.~S.,  1999, \mn@doi [\apjl] {10.1086/311942}, 514, L87

\bibitem[\protect\citeauthoryear{{Younger}, {Cox}, {Seth}  \&
  {Hernquist}}{{Younger} et~al.}{2007}]{younger2007}
{Younger} J.~D.,  {Cox} T.~J.,  {Seth} A.~C.,   {Hernquist} L.,  2007, \mn@doi
  [\apj] {10.1086/521976}, 670, 269

\bibitem[\protect\citeauthoryear{{Zabludoff}, {Zaritsky}, {Lin}, {Tucker},
  {Hashimoto}, {Shectman}, {Oemler}  \& {Kirshner}}{{Zabludoff}
  et~al.}{1996}]{zabludoff1996}
{Zabludoff} A.~I.,  {Zaritsky} D.,  {Lin} H.,  {Tucker} D.,  {Hashimoto} Y.,
  {Shectman} S.~A.,  {Oemler} A.,   {Kirshner} R.~P.,  1996, \mn@doi [\apj]
  {10.1086/177495}, 466, 104

\bibitem[\protect\citeauthoryear{{van Dokkum}}{{van
  Dokkum}}{2005}]{vandokkum2005}
{van Dokkum} P.~G.,  2005, \mn@doi [\aj] {10.1086/497593}, 130, 2647

\bibitem[\protect\citeauthoryear{{van der Wel}, {Holden}, {Zirm}, {Franx},
  {Rettura}, {Illingworth}  \& {Ford}}{{van der Wel}
  et~al.}{2008}]{vanderwel2008a}
{van der Wel} A.,  {Holden} B.~P.,  {Zirm} A.~W.,  {Franx} M.,  {Rettura} A.,
  {Illingworth} G.~D.,   {Ford} H.~C.,  2008, \mn@doi [\apj] {10.1086/592267},
  688, 48

\bibitem[\protect\citeauthoryear{{van der Wel} et~al.,}{{van der Wel}
  et~al.}{2012}]{vanderwel2012}
{van der Wel} A.,  et~al., 2012, \mn@doi [\apjs] {10.1088/0067-0049/203/2/24},
  203, 24

\bibitem[\protect\citeauthoryear{{van der Wel} et~al.,}{{van der Wel}
  et~al.}{2014}]{vanderwel2014}
{van der Wel} A.,  et~al., 2014, \mn@doi [\apj] {10.1088/0004-637X/788/1/28},
  788, 28

\makeatother
\end{thebibliography}


\bsp	
\label{lastpage}
\end{document}